\documentclass[showpacs,aps,prd,longbibliography,nofootinbib,showkeys,superscriptaddress,twocolumn,longbibliography]{revtex4-2}
\usepackage{graphicx}
\usepackage{bm}
\usepackage{multirow}
\usepackage{natbib}
\usepackage{array}
\usepackage{amsmath}
\usepackage{mathrsfs}
\usepackage{amssymb}
\usepackage{float}
\usepackage{xcolor}
\usepackage{enumerate}
\usepackage{color}
\usepackage{slashed}
\usepackage{hyperref}
\usepackage{url}
\usepackage{txfonts}
\usepackage{mathtools}
\usepackage{booktabs}
\usepackage{BOONDOX-cal}
\usepackage{orcidlink}
\usepackage{longtable}
\usepackage{makecell}
\usepackage{cancel}

\allowdisplaybreaks[4]
\hypersetup{hypertex=true,colorlinks=true,linkcolor=blue,anchorcolor=red,citecolor=blue}

\newcommand{\dif}{\mathrm{d}}
\newcommand{\e}{\mathrm{e}}

\begin{document}
\title{Rapidity scan approach for net-baryon cumulants with a statistical thermal model}
\author{Jianing Li\orcidlink{0000-0001-7193-7237}}\email{lijianing@impcas.ac.cn}
\affiliation{Southern Center for Nuclear-Science Theory (SCNT), Institute of Modern Physics, Chinese Academy of Sciences, Huizhou, Guangdong 516000, China}
\author{Lipei Du\orcidlink{0000-0002-3029-6602}}\email{lipei.du@mail.mcgill.ca}
\affiliation{Department of Physics, McGill University, Montreal, Quebec H3A 2T8, Canada}
\author{Shuzhe Shi\orcidlink{0000-0002-3042-3093}}\email{shuzhe-shi@tsinghua.edu.cn}
\affiliation{Department of Physics, Tsinghua University, Beijing 100084, China}
\date{\today}
\begin{abstract}
 Utilizing rapidity-dependent measurements to map the QCD phase diagram provides a complementary approach to traditional beam-energy-dependent measurements around midrapidity. The changing nature of thermodynamic properties of QCD matter along the beam axis in heavy-ion collisions at low collision energies both motivates and poses challenges for this method. In this study, we derive the analytical cumulant-generating function for subsystems within distinct rapidity windows, while accounting for global net-baryon charge conservation of the full system. Rapidity-dependent net-baryon cumulants are then calculated for a system exhibiting inhomogeneity along the beam axis, and their sensitivity to finite acceptances through changing rapidity bin widths is explored. We highlight the nontrivial behaviors exhibited by these cumulants, underscoring their importance in establishing a noncritical baseline for interpreting net-proton cumulants in the search for the QCD critical point. Finally, we discuss the implications of the rapidity scan for mapping the QCD phase diagram within the current context.
\end{abstract}
\maketitle
\section{Introduction}\label{Sec1}

One of the fundamental goals in nuclear physics is to achieve a quantitative understanding of the phase structure within the quantum chromodynamics (QCD) phase diagram~\cite{An:2021wof,Du:2024wjm}. At vanishing baryon chemical potential $\left(\mu_B=0\right)$, extensive first-principle calculations of lattice QCD have yielded compelling evidence for a smooth crossover between the quark-gluon plasma (QGP) phase and the hadron resonance gas phase~\cite{Aoki:2006we}. Interestingly, with a hybrid model, Bayesian analysis using various experimental measurements results in an equation of state (EoS) consistent with the lattice QCD EoS~\cite{Pratt:2015zsa}. Recent experimental measurements of net-proton cumulants from high-energy nuclear collisions at the top energy available at the BNL Relativistic Heavy Ion Collider (RHIC) also suggest such a smooth phase transition when $\mu_B$ is small~\cite{STAR:2021rls}. In the region of high $\mu_B$, effective QCD models postulate the presence of a first-order transition line that ends at a QCD critical point~\cite{Bowman:2008kc,Ejiri:2008xt}. Regrettably, the formidable sign problem, a persistent issue within lattice QCD~\cite{Barbour:1997ej}, hinders our ability to provide reliable predictions at nonzero baryon chemical potentials regarding the location of the critical point through first-principle calculations. The primary approach for seeking the QCD critical point involves the experimental measurement of the cumulants of net-baryon charge~\cite{ALICE:2019nbs,STAR:2020tga,HADES:2020wpc,STAR:2021iop,STAR:2022vlo}, as they are believed to exhibit sensitivity to the correlation length~\cite{Stephanov:1999zu,Stephanov:2008qz,Stephanov:2011pb,Asakawa:2009aj}, which undergoes a divergence at the critical point. Moreover, the cumulants of the net-baryon charge may also encode information about the EoS of QCD matter produced in heavy-ion collisions~\cite{Sorensen:2020ygf,Sorensen:2021zme,Sorensen:2023zkk}.

In heavy-ion collisions, the reduction of the center-of-mass energy ($\sqrt{s_{NN}}$) leads to an increased number of net-baryon charges within the collision zone, thereby allowing for the creation of systems at higher $\mu_B$~\cite{Braun-Munzinger:2007edi}. This rationale underpins the method of {\em beam energy scan}, where collisions are systematically conducted across a range of high to low beam energies to scan the QCD phase diagram. If the critical point exists within the scanned region, it is expected that the correlation length and, consequently, the cumulants of net-baryon charge will exhibit a nonmonotonic behavior as a function of beam energy~\cite{ALICE:2019nbs,STAR:2020tga,HADES:2020wpc,STAR:2021iop,STAR:2022vlo}. Recent measurements of net-proton cumulants conducted by STAR Collaboration~\cite{STAR:2020tga,STAR:2021iop,STAR:2022vlo} during the first phase of the Beam Energy Scan (BES-I) study indicate the presence of this nonmonotonic behavior, albeit with notable statistical uncertainties. Besides varying beam energy, investigating different rapidity windows provides another means to explore medium properties across varying $\mu_B$~\cite{Begun:2018efg,Brewer:2018abr,Karpenko:2018xam,Du:2021zqz,Du:2021uxo}. This approach, known as {\em rapidity scan}, gains particular significance at low beam energies, where the medium undergoes substantial changes along the beam axis. Along this line, researchers have explored potential critical signatures by analyzing rapidity-dependent cumulants~\cite{Brewer:2018abr, Yin:2018ejt,Barej:2023xqt}.

Mapping the QCD phase diagram through experimental measurements at varying beam energies or within different rapidity windows is a complex undertaking. This often necessitates model-to-data comparisons, wherein the models employed can range from multistage dynamical models to statistical thermal models. The focus of this study is on the latter approach, which generally assumes that strongly interacting QCD matter attains thermal and chemical equilibrium at hadronization and models it as a hadron resonance gas (statistical thermal)~\cite{Karsch:2003zq,Karsch:2003vd,Allton:2005gk, Huovinen:2009yb,Majumder:2010ik, Ratti:2010kj, Vovchenko:2019pjl}. The implementation of statistical thermal models has achieved great success in interpreting particle abundances~\cite{Becattini:2003wp,Tawfik:2004sw, Cleymans:2005xv, Andronic:2008gu, Andronic:2017pug, Du:2022yok,Du:2023efk}, enabling the extraction of temperature and baryon chemical potential at chemical freeze-out~\cite{Braun-Munzinger:2003pwq,Wheaton:2004qb, Ejiri:2005wq,Karsch:2010ck,Alba:2014eba, Tawfik:2014eba, STAR:2017sal,STAR:2017sal}. These models have also been extended to describe cumulants of fluctuating particle multiplicity, demonstrating good agreement with both lattice QCD results and experimental measurements~\cite{Karsch:2010ck, Braun-Munzinger:2011shf, Borsanyi:2018grb, Karsch:2016yzt, Braun-Munzinger:2020jbk, Vovchenko:2020kwg, Vovchenko:2021kxx, Vovchenko:2021yen}. Given this success, statistical thermal models serve as a noncritical baseline for net-proton cumulants measured at various beam energies, particularly above $\sqrt{s_{NN}}=27$ GeV, where they align with experimental observations after accounting for net-baryon conservation and interactions~\cite{Vovchenko:2021kxx, Braun-Munzinger:2020jbk}. Assessing whether the deviations observed below this beam energy are indicative of criticality holds significant importance~\cite{STAR:2020tga, STAR:2021iop}.

While there are remarkable agreements between measurements and models, it is essential to note that, to increase statistical precision, identified particle yields and high-order cumulants are typically measured within finite rapidity windows centered around midrapidity (e.g., net-proton cumulants within $|y|<1$ by STAR), spanning different beam energies. However, results from statistical thermal models and lattice QCD are generally derived with specific values for temperature and chemical potential applied. As previously mentioned, the thermodynamic properties of the system exhibit more pronounced variations along rapidity at lower beam energies. This suggests that observables measured within finite rapidity windows result from averaging over thermodynamic properties that change more dramatically at lower beam energies. In essence, the extent of this averaging varies across different beam energies, introducing complexity in efforts to link the nonmonotonic behavior in net-proton cumulants and their deviation from the noncritical statistical thermal baseline within $7.7 \leq \sqrt{s_{NN}} \leq 27~\text{GeV}$~\cite{STAR:2020tga, STAR:2021iop} to potential critical phenomena.

The upcoming measurements from BES-II, featuring a wider range of beam energies and improved statistical precision, are expected to play a crucial role in elucidating the nonmonotonic trend in high-order cumulants, potentially offering insights into the critical nature of the phenomenon. Additionally, BES-II will also expand the availability of rapidity-dependent measurements. To enhance the comprehension of these forthcoming and exciting measurements, in this study, we investigate the statistical thermal model while incorporating rapidity-dependent freeze-out profiles to account for the variances in thermodynamic properties along the beam axis. Our analysis focuses on the exploration of higher-order net-baryon cumulants within finite rapidity windows while taking into account global baryon conservation. Notably, we expand our analysis beyond midrapidity and consider various rapidity windows, aligning with the spirit of a rapidity scan approach. Our investigation serves as a valuable step in mapping the QCD phase diagram through rapidity-dependent measurements. Furthermore, it offers a means to establish a noncritical baseline for rapidity-dependent cumulants, aiding in the search for the QCD critical point.

This paper is organized as follows: Section~\ref{Sec2} provides a concise overview of two thermal model scenarios considered in this study: the single-source model and the continuous-source model. In Sec.~\ref{Sec3}, we present the derivation of the cumulant-generating function for an inhomogeneous system in the beam axis, employing the statistical thermal model while considering global charge conservation. Section~\ref{Sec4} is dedicated to the calculation of rapidity-dependent net-baryon cumulants at $\sqrt{s_{NN}}=19.6$ GeV, accompanied by a discussion on the rapidity scan of the QCD phase diagram utilizing these cumulants. Finally, Sec.~\ref{Sec5} offers a summary.
\section{Two thermal model scenarios}\label{Sec2}

\begin{figure}[!tpb]
    \centering
    {\includegraphics[width=0.4\textwidth]{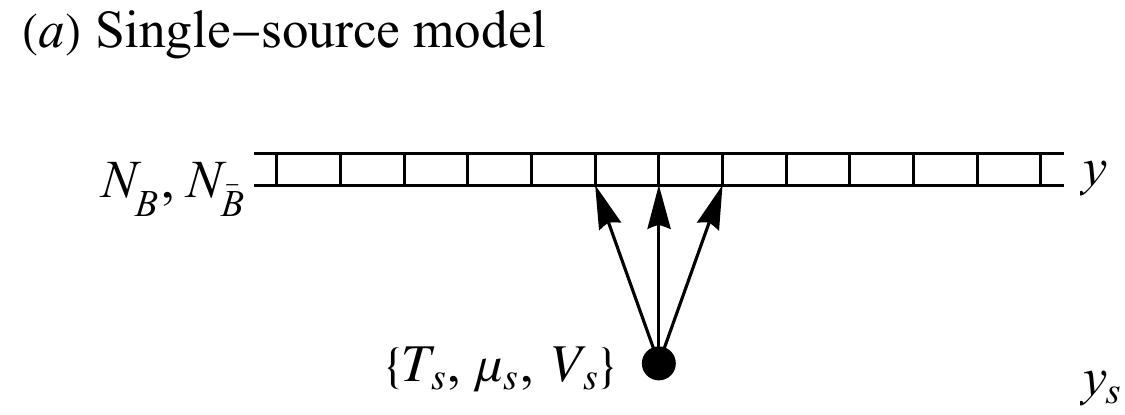}}
    {\includegraphics[width=0.4\textwidth]{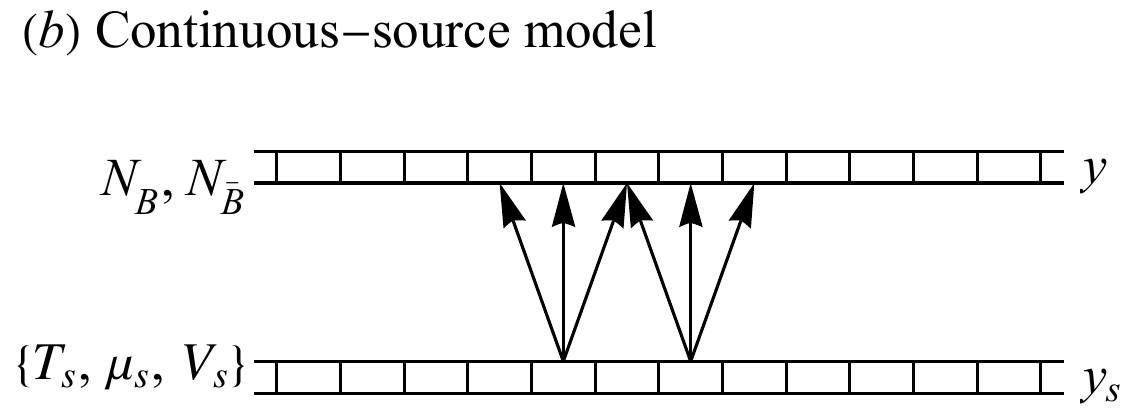}}
    \caption{ Sketches of ($a$) the single-source model and ($b$) the continuous-source model. The band or dot at the bottom represents the thermal source(s), while the band on top illustrates the radiated particles, along rapidity. $y_s$ and $y$ denote the rapidity of the thermal source and the rapidity windows of the emitted particles, respectively. $T_s$, $\mu_s$, and $V_s$ correspond to the temperature, baryon chemical potential, and volume of the thermal source indexed by $s$. $N_B$ and $N_{\bar{B}}$ indicate the numbers of baryons and antibaryons within a given rapidity window.}
    \label{fig1}
\end{figure}

In statistical thermal modeling applied to heavy-ion collisions, both the single-source model and the continuous-source model are implemented for interpreting measured particle distributions. A detailed visualization of these models is provided in FIG.~\ref{fig1} to aid in better understanding.

As illustrated in FIG.~\ref{fig1}, in the \emph{single-source model}, all particles from a collision are assumed to originate from a single point-like source, simplifying the radiation process with a single set of thermodynamic parameters~\cite{Braun-Munzinger:1999hun, Wheaton:2004qb, Begun:2018efg}. This source, often represented as a Dirac $\delta$ function in rapidity, has a finite volume in spacetime coordinate with uniform, static temperature, and baryon chemical potential. While effective for describing some thermal characteristics and calculating observables such as particle yields at a specific rapidity, this model may oversimplify actual experimental dynamics. The \emph{continuous-source model} expands upon the single-source model by viewing the system as a superposition of thermal sources distributed continuously in rapidity space. Unlike the single-source model, it recognizes that particles measured in different rapidity bins can originate from various regions or sources rather than a singular one. This model allows for the consideration of a continuous distribution of radiation sources, each potentially characterized by distinct thermodynamic conditions or emission properties. Consequently, it offers a more detailed representation of the radiation mechanism, including effects like thermal smearing and longitudinal flows~\cite{Du:2023gnv, Gao:2023xiv}. The selection between these models depends on the desired level of accuracy for describing both experimental observations and theoretical predictions.
\section{Net-baryon cumulants of an inhomogeneous system}\label{Sec3}

In the quest for the elusive QCD critical point in heavy-ion collision experiments, higher-order cumulants of conserved charges play a crucial role due to their sensitivity to the correlation length in a QCD medium. 
While each collision fireball is a closed system with conserved charges, the cumulants of net baryon fluctuation become nontrivial when considering the particles within the kinematic region that can be detected by the detectors. This finite acceptance significantly impacts the measured charge cumulants, emphasizing the need for caution when exploring critical phenomena through cumulant measurements~\cite{Braun-Munzinger:2020jbk}. In this study, we calculate the net-charge cumulants for a subsystem within a detected region, as part of the entire collision fireball that is a closed system where net-charge is conserved. In particular, the detected region may correspond to different rapidity windows, enabling the investigation of rapidity-dependent cumulants.

Adopting the textbook approach that calculates the cumulants of a net charge within the detected region, we express the first six orders of cumulants $\kappa_n^B~\left(n=1,\,2,\,\cdots, 6\right)$ for the net-baryon number within the acceptance ($B_A$, with subscript $A$ for ``accepted'') as~\cite{Luo:2017faz}
\begin{equation}
	\begin{split}
		\kappa_1^B&=\left<B_A\right>\,,\\
		\kappa_2^B&=\left<\left(\delta B_A\right)^2\right>\,,\\
		\kappa_3^B&=\left<\left(\delta B_A\right)^3\right>\,,\\
		\kappa_4^B&=\left<\left(\delta B_A\right)^4\right>-3\left<\left(\delta B_A\right)^2\right>^2\,,\\
		\kappa_5^B&=\left<\left(\delta B_A\right)^5\right>-10\left<\left(\delta B_A\right)^3\right>\left<\left(\delta B_A\right)^2\right>\,,\\
		\kappa_6^B&=\left<\left(\delta B_A\right)^6\right>-15\left<\left(\delta B_A\right)^4\right>\left<\left(\delta B_A\right)^2\right>\\
		&\quad-10\left<\left(\delta B_A\right)^3\right>^2+30\left<\left(\delta B_A\right)^2\right>^3\,.
	\end{split}
\end{equation}
Here, $\left<\cdots\right>$ denotes the ensemble average, and $\delta B_A\equiv B_A-\left<B_A\right>$ represents the event-by-event fluctuation of $B_A$. The cumulants defined above can be systematically obtained by taking derivatives of the net-baryon cumulant-generating function $g_B(t)$~\cite{kardar_2007},
\begin{equation}
	\kappa_n^B = \frac{\mathrm{d}^n g_B(t)}{\mathrm{d} t^n}\bigg|_{t=0}\,,
 \label{eq:cumulants}
\end{equation}
which is defined as follows:
\begin{equation}
	g_B\left(t\right)=\ln\left[\sum_{B_A=-\infty}^{\infty}{\e}^{t B_A}P\left(B_A\right)\right]\,,
 \label{eq:gB}
\end{equation}
where $P\left(B_A\right)$ represents the probability of detecting an event with net-baryon number $B_A$ within the accepted region.

It is essential to highlight that,  during the evolution, the total net-baryon number $B$ deposited into the collision zone by the participants is conserved, and thus the canonical ensemble (CE) should be used to address the $B$ in the full space of the entire system. Nevertheless, in principle, the total net-baryon number $B$ deposited into the collision zone can fluctuate on an event-by-event basis, influenced mainly by impact parameter, and other second-order eﬀects such as volume ﬂuctuations within a centrality class. Hence, $P\left(B_A\right)$ should be considered as a conditional probability distribution $P\left(B_A|B\right)$, denoting the probability of detecting a net-baryon number $B_A$ within the accepted region when the full system exhibits a total net-baryon number $B$. However, since the centrality fluctuations are typically mitigated by experimentalists in the measurements, we do not consider the fluctuations of $B$ in this analysis. On the other hand, the measured net-baryon number $B_A$ within the acceptance is subject to fluctuations. These measured baryons are emitted from subsystems, capable of particle exchange with neighboring subsystems, necessitating their treatment as open systems. Consequently, we treat each subsystem as a grand canonical ensemble (GCE) system within a full system treated as CE. This approach aligns with the treatment implemented by the subensemble acceptance method (SAM) framework~\cite{Vovchenko:2020kwg,Vovchenko:2020tsr}. 

According to Eqs.~\eqref{eq:cumulants} and \eqref{eq:gB}, computing cumulants $\kappa_n^B$ from the generating function $g_B(t)$ requires obtaining $P\left(B_A|B\right)$. In events characterized by the probability distribution $P\left(B_A|B\right)$, if we assume that $N_A$ baryons and $\bar{N}_A$ anti-baryons are observed in the acceptance region, both subject to fluctuations, then they are constrained by $N_A-\bar{N}_A = B_A$. Notably, outside the accepted region, there exists net-baryon number of $B_R = B - B_A$, with the subscript $R$ denoting ``rejected.'' If we denote the number of baryons and anti-baryons created but falling outside the accepted region as $N_R$ and $\bar{N}_R$, respectively, then they are constrained by $N_R-\bar{N}_R = B_R = B - B_A$.
Thus, the probability $P\left(B_A|B\right)$ is the sum of the probabilities of all the events that satisfy those constraints,
\begin{equation}
\begin{split}
	P\left(B_A|B\right) =\;& \sum_{N_A, \bar{N}_A, N_R, \bar{N}_R = 0}^{\infty}
        \delta_{B_A, N_A-\bar{N}_{A}} \delta_{B-B_A, N_R-\bar{N}_{R}}
    \\&\qquad\times
        P_A\left(N_A\right) P_A\left(\bar{N}_A\right) 
        P_R\left(N_R\right) P_R\left(\bar{N}_R\right)
    \,.
\end{split}\label{eq:PB.1}
\end{equation}
Here, $P_A(n)$ and $P_R(n)$ stand as independent probability distributions. The former signifies the probability of having $n$ particles (or antiparticles) accepted within the kinematic region of interest, while the latter represents the probability of $n$ particles (or antiparticles) being rejected.
 
Observing that the measured particle (or antiparticle) numbers stem from various hadron species indexed by $h$, which, in turn, originate from distinct thermal sources indexed by $s$, the probabilities in Eq.~\eqref{eq:PB.1} can be further broken down as follows:
\begin{align}
    P_i(n_i) = \left\{\prod_{s,h}\sum_{n_{i,h}^s=0}^{+\infty}\right\} \prod_{s,h} P_i\left(n_{i,h}^s\right) \delta_{n_i,\sum_{h,s}n_{i,h}^s}\,,
\label{eq:Pi.1}
\end{align}
where $i{\,=\,}A,\,R$ and $n{\,=\,}N,\,\bar{N}$. As an example, ${N}_{A,h}^s$ ($\bar{N}_{A,h}^s$) denotes the particle (antiparticle) numbers within the accepted region carried by hadron species $h$ radiated from the thermal source $s$. Note that the numbers of hadrons ${N}_{i,h}$ and their corresponding  antiparticles $\bar{N}_{i,h}$ are distinguished explicitly, and here $h$ labels the species. The remaining task involves calculating $P_i\left(n_{i,h}^s\right)$ in Eq.~\eqref{eq:Pi.1}. In statistical physics, the probability can be calculated from the single-particle partition function. We shall derive the single-particle partition function of a GCE with a finite baryon chemical potential from a CE.\footnote{In this section, we will introduce various types of single-particle partition functions. For readers seeking a quick reference, we summarize their definitions and meanings in TABLE~\ref{Conventions} in Appendix~\ref{AppA}.} 

We treat the rapidity distribution of the thermal source indexed by $s$ as a Dirac $\delta$ function centered at the rapidity of the source $y_s \equiv \ln\left[\left(1+v_z^s\right)/\left(1-v_z^s\right)\right]/2$, where $v_z^s$ is the velocity of source $s$ along the beam axis. For an infinitesimal thermal source at midrapidity $y_s=0$, characterized by a volume element $\dif{V_s}$ and a uniform temperature $T_s$, its CE single-particle partition function for $h$ is given by
\begin{equation}
	\label{eq:zh1}
	\begin{split}
	z_h^s\equiv z_h\left(T_s,{\dif}V_s\right) &=\frac{d_h}{\left(2\pi\right)^3}\left(\int{\e}^{-\frac{\sqrt{m_h^2+\boldsymbol{p}^2}}{T_s}}{\dif}^3\boldsymbol{p}\right)\dif{V_s}\\
	&=\frac{d_h}{2\pi^2}T_sm_h^2K_2\left(\frac{m_h}{T_s}\right)\dif{V_s}\,,
	\end{split}
\end{equation}
where the Maxwell--Boltzmann approximation has been assumed and $K_\nu\left(x\right)$ represents the modified Bessel function of the second kind. Here, $\boldsymbol{p}$, $d_h$, and $m_h$ denote the momentum, degeneracy factor, and mass of $h$, respectively. For brevity, all dependencies on the properties of thermal source $s$, including $y_s$, $T_s$, ${\dif}V_s$, and the forthcoming baryon chemical potential $\mu_s$, are abbreviated using the superscript ``$s$" in $z_h^s$.
 
Denoting the rapidity of a radiated hadron as $y\equiv\ln\left[\left(1+v_z\right)/\left(1-v_z\right)\right]/2$, where $v_z$ represents its velocity along the beam axis, the integral measure in Eq.~\eqref{eq:zh1} can be reformulated as ${\dif}^3\boldsymbol{p}= p_\mathrm{T}\sqrt{m_h^2+p_\mathrm{T}^2} \cosh y\,{\dif}p_\mathrm{T} {\dif} \phi{\dif} y$, with $p_\mathrm{T}$ being the transverse momentum and $\phi$ the azimuthal angle of the hadron. The rapidity-differential single-particle partition function is given by
\begin{equation}\label{eq:zhs1}
    \begin{split}
        z_h^s\left(y\right)&\equiv z_h \left(T_s,{\dif}V_s;y\right)\\
        &=\frac{d_h}{\left(2\pi\right)^3}\dif{V_s}\int_0^{2\pi}{\dif}\phi\int_0^{+\infty}p_\mathrm{T}\,{\dif}p_\mathrm{T}\\
        &\times\sqrt{m_h^2+p_\mathrm{T}^2} \cosh y\,{\e}^{-\frac{\sqrt{m_h^2+p_\mathrm{T}^2}}{T_s}\cosh y}\\
	&=\frac{d_hT_s}{\left(2\pi\right)^2}\left(m_h^2 + \frac{2m_h T_s}{\cosh y}+ \frac{2T_s^2}{\cosh^2 y}\right){\e}^{-\frac{m_h \cosh y}{T_s}}\dif{V_s}\,.
    \end{split}
\end{equation}
The partition function of a thermal source at a nonzero rapidity $y_s$ can be acquired by a straightforward boost of Eq. \eqref{eq:zhs1}, expressed as $z_h^s(y-y_s)$ in general.

For a GCE with a finite baryon chemical potential ($\mu_s$), the single-particle partition functions for baryons and anti-baryons can be respectively expressed as~\cite{Cleymans:1991wr, Ko:2000vp, Braun-Munzinger:2003pwq, Cleymans:2004iu, Braun-Munzinger:2011xux, Braun-Munzinger:2011shf}
\footnote{On the right-hand side, $z_h^s$ in normal font represents the CE, while $\mathcal{z}_h^s$ in calligraphic font on the left-hand side signifies the GCE.}
\begin{equation}\label{eq:zhs2}
	\begin{gathered}
	\mathcal{z}_h^s\left(y-y_s\right)={\e}^{B_h\frac{\mu_s}{T_s}}z_h^s\left(y-y_s\right)\,,\\
	\bar{\mathcal{z}}_h^s\left(y-y_s\right)={\e}^{-B_h\frac{\mu_s}{T_s}}z_h^s\left(y-y_s\right)\,,
	\end{gathered}
\end{equation}
where $B_h$ is positive, representing the (absolute) baryon number carried by both the baryon and its antiparticle indexed by $h$.
Consequently, the expectation values of the baryon and the antibaryon numbers at rapidity $y$, radiated from the thermal source at $y_s$, are given by, respectively,
\begin{equation}
	\label{baryon number}
	\begin{gathered}
		\left<N_h^s\right>\left(y-y_s\right)=
        B_h \mathcal{z}_h^s\left(y-y_s\right)\,,\\
		\left<\bar{N}_h^s\right>\left(y-y_s\right)=
        B_h \bar{\mathcal{z}}_h^s\left(y-y_s\right)\,.
	\end{gathered}
\end{equation}
Considering the accepted region characterized by a rapidity window centered at $y_c$ with a width $w$, denoted as $y \in \left[y_c - w/2,\, y_c + w/2\right]$, the GCE single-particle partition function of source $s$ can be segregated into a segment within the acceptance window,
\begin{equation}\label{eq:zAhs1}
	\begin{gathered}
	\mathcal{z}_{A,h}^s=\int_{y_c - \frac{w}{2}}^{y_c + \frac{w}{2}}\,\mathcal{z}_h^s\left(y-y_s\right)\,{\dif}y\,,\\
	\bar{\mathcal{z}}_{A,h}^s=\int_{y_c - \frac{w}{2}}^{y_c + \frac{w}{2}}\,\bar{\mathcal{z}}_h^s\left(y-y_s\right)\,{\dif}y\,,
	\end{gathered}
\end{equation} 
and another outside it (i.e., ``rejected''),
\begin{equation}\label{eq:zRhs1}
	\begin{gathered}
		\mathcal{z}_{R,h}^s=\int_{-\infty}^{+\infty}\,\mathcal{z}_h^s\left(y-y_s\right)\,{\dif}y-\mathcal{z}_{A,h}^s\,,\\
	   \bar{\mathcal{z}}_{R,h}^s=\int_{-\infty}^{+\infty}\bar{\mathcal{z}}_h^s\left(y-y_s\right)\,{\dif}y-\bar{\mathcal{z}}_{A,h}^s\,.
	\end{gathered}
\end{equation} 

Now, we proceed with the computation of the probability associated with observing $N_{A,h}^s$ ($N_{R,h}^s$) hadrons emitted from source $s$ within (outside) the acceptance range in Eq. \eqref{eq:Pi.1} and obtain~\cite{Pathria1972}
\begin{equation}\label{eq:Pi.2}
    P_i\left(N_{i,h}^s\right) = \frac{\left(\mathcal{z}_{i,h}^s\right)^{N_{i,h}^s}}{N_{i,h}^s!} \e^{-\mathcal{z}_{i,h}^s}\,,\qquad i=A,\,R\,.
\end{equation}
Here we have used the Poisson distribution to describe these independent events and to determine the probability for the anti-particles, simply substitute $\mathcal{z}_{i,h}^s$ with $\bar{\mathcal{z}}_{i,h}^s$ and $N_{i,h}^s$ with $\bar{N}_{i,h}^s$. Substituting Eq. \eqref{eq:Pi.2} into Eq. \eqref{eq:Pi.1}, we find
\begin{equation}\label{eq:Pi.3}
    \begin{split}
    P_i(n_i) &= \left\{\prod_{s,h}\sum_{n_{i,h}^s=0}^{+\infty}\right\} \prod_{s,h}   \frac{\left(\mathcal{z}_{i,h}^s\right)^{n_{i,h}^s}}{n_{i,h}^s!} \e^{-\mathcal{z}_{i,h}^s} \delta_{n_i,\sum_{h,s}n_{i,h}^s}
\\&=
    \frac{\left(\sum_{s,h}\mathcal{z}_{i,h}^s\right)^{n_i}}{n_i!}
    \e^{-\sum_{s,h}\mathcal{z}_{i,h}^s}
\equiv
    \frac{\mathcal{z}_i^{n_i}}{n_i!} \e^{-\mathcal{z}_i}\,,
    \end{split}
\end{equation}
where we have introduced partition functions summed over different thermal sources indexed by $s$ and various baryon species indexed by $h$:
\begin{equation}\label{eq:z2}
		\mathcal{z}_i
            =\sum_{s,h}\mathcal{z}_{i,h}^s\,,\quad
		\bar{\mathcal{z}}_i
            =\sum_{s,h}\bar{\mathcal{z}}_{i,h}^s\,,\\
	\quad i=A,\,R.
\end{equation}
For thermal sources distributed continuously across $y_s$, the summation over $s$ is replaced by integration, $\sum_{s} [\cdots] \dif V_s \rightarrow \int_s[\cdots]\left(\dif V_s/\dif y_s\right)\dif y_s$, under the continuum limit.

Advancing towards Eq.~\eqref{eq:PB.1}, we proceed to compute the subsequent expression involving summation over $N_A$ and $\bar{N}_A$,
\begin{equation}
\begin{split}
    &\quad\sum_{N_A, \bar{N}_A = 0}^{\infty}
    \delta_{B_A, N_A-\bar{N}_{A}} 
    P_A\left(N_A\right) P_A\left(\bar{N}_A\right)
    \\
    &=\e^{-{\mathcal{z}}_A-\bar{\mathcal{z}}_A} 
    \left(\frac{\mathcal{z}_{A}}{\bar{\mathcal{z}}_A}\right)^\frac{B_A}{2}
    I_{B_A}\left(2\sqrt{\mathcal{z}_{A} \bar{\mathcal{z}}_{A}}\right)\,,
\end{split}
\end{equation}
and upon substituting $A$ with $R$, the corresponding result for the particles outside the acceptance range is obtained. Here, $I_n\left(x\right)$ is the modified Bessel function of the first kind. Ultimately, the probability described in Eq.~\eqref{eq:PB.1} can be evaluated as follows: 
\begin{equation}\label{eq:PB}
\begin{split}
    P\left(B_A|B\right)&\propto\left(\frac{\mathcal{z}_{A}}{\bar{\mathcal{z}}_A}\right)^\frac{B_A}{2}I_{B_A}\left(2\sqrt{\mathcal{z}_{A}\bar{\mathcal{z}}_{A}}\right)\\
    &\quad\times\left(\frac{\mathcal{z}_{R}}{\bar{\mathcal{z}}_{R}}\right)^\frac{B-B_A}{2} I_{B-B_A}\left(2\sqrt{\mathcal{z}_{R}\bar{\mathcal{z}}_{R}}\right)\,,
\end{split}
\end{equation}
up to a normalization factor. By substituting Eq.~\eqref{eq:PB} into Eq.~\eqref{eq:gB} and utilizing Graf's addition formula~\cite{Bateman:1953higher,Bzdak:2012an,Braun-Munzinger:2020jbk},
\begin{equation}
    \begin{split}
        &\quad\sum_kw^kI_k\left(x\right)I_{n-k}\left(y\right)\\
        &=\left(w\frac{y+wx}{x+wy}\right)^\frac{n}{2}I_n\left(\sqrt{x^2+y^2+\frac{1+w^2}{w}xy}\right)\,,
    \end{split}
\end{equation}
we derive a straightforward and exact formula for the net-baryon (accepted) cumulant-generating function,
\begin{equation}\label{eq:gB1}
    \begin{split}
	g_B\left(t\right)&=\frac{B}{2}\left\{\ln\left[q_1\left(t\right)\right]-\ln\left[q_2\left(t\right)\right]\right\}\\
	&\quad+\ln\left\{I_B\left[2\sqrt{\mathcal{z}\bar{\mathcal{z}}}\sqrt{q_1\left(t\right)q_2\left(t\right)}\right]\right\}+\text{const}\,,
    \end{split}
\end{equation}
where
\begin{equation}
	\label{eq:gB2}
	\begin{split}
		q_1\left(t\right) =\,& 1-\alpha_B+\alpha_B\,\e^t\,,\quad
        \mathcal{z} = \mathcal{z}_A + \mathcal{z}_R\,,\quad
	    \alpha_{B} = \frac{\mathcal{z}_{A}}{\mathcal{z}}\,,
\\
		q_2\left(t\right) =\,& 1-\alpha_{\bar{B}}+\alpha_{\bar{B}}\,\e^{-t}\,,\quad
        \bar{\mathcal{z}}= \bar{\mathcal{z}}_A + \bar{\mathcal{z}}_R\,,\quad
	       \alpha_{\bar{B}} = \frac{\bar{\mathcal{z}}_{A}}{\bar{\mathcal{z}}}\,.
	\end{split}
\end{equation}
Equations (\ref{eq:gB1}) and (\ref{eq:gB2}) reveal that the cumulant generating function $g_B(t)$, along with all the cumulants $\kappa^B_n$, are functions of $\alpha_B$, $\alpha_{\bar{B}}$, and $\sqrt{\mathcal{z} \bar{\mathcal{z}}}$. In Eqs.~\eqref{eq:gB1} and~\eqref{eq:gB2}, ${\mathcal{z}}$ and $\bar{\mathcal{z}}$ represent all particles, irrespective of whether they fall inside or outside the rapidity acceptance range. On the other hand, $\alpha_{{B}}$ and $\alpha_{\bar{B}}$ specifically pertain to particles within the acceptance range defined by $y \in [y_c - w/2, y_c + w/2]$.

Several insights can be drawn from the results. First, if we exclusively consider baryons carrying only one baryon charge and exclude light nuclei with $B_h>1$, $\mathcal{z}$ and $\bar{\mathcal{z}}$ yield the expected values of baryons $\left<N_B\right>$ and antibaryons $\left<N_{\bar{B}}\right>$ in the full phase space. This relationship is evident by substituting Eq.~\eqref{eq:z2} into Eq.~\eqref{baryon number}. Similarly, $\mathcal{z}_A$ and $\bar{\mathcal{z}}_A$ represent the accepted baryon and antibaryon numbers, denoted as $\left<N_B\right>_A$ and $\left<N_{\bar{B}}\right>_A$, respectively. Consequently, $\alpha_B=\left<N_B\right>_A/\left<N_B\right>$ and $\alpha_{\bar{B}}=\left<N_{\bar{B}}\right>_A/\left<N_{\bar{B}}\right>$ can be interpreted as the acceptances for baryons and anti-baryons within a kinematic region of interest. Second, in Eq.~\eqref{eq:gB1}, the first line represents fluctuations attributed solely to either baryons or antibaryons, while the second line reflects fluctuations influenced by both baryons and antibaryons. Moreover, both $q_1\left(t\right)$ and $q_2\left(t\right)$ are bound by the global net-baryon number conservation condition $B=\left<N_B\right>-\left<N_{\bar{B}}\right>$, stemming from the imposed constraints of the CE. Third, it should be emphasized that our results of ${\mathcal{z}}_A$, ${\mathcal{z}}_R$, $\bar{\mathcal{z}}_A$, and $\bar{\mathcal{z}}_R$, and subsequently $\alpha_B$, $\alpha_{\bar{B}}$, and $\sqrt{\mathcal{z} \bar{\mathcal{z}}}$, are computed as summations across all thermal sources. Equations~\eqref{eq:gB1} and~\eqref{eq:gB2} would yield results consistent with those obtained for a singular or an extended but homogeneous source, as outlined in Eq. (A.1) of Ref.~\cite{Braun-Munzinger:2020jbk}, quoted below:
\begin{equation}
    \label{eq:gB3}
	\begin{split}
		g_B\left(t\right)&=\frac{B}{2}\left\{\ln\left[q_1\left(t\right)\right]-\ln\left[q_2\left(t\right)\right]\right\}\\
		&\quad+\ln\left\{I_B\left[2z\sqrt{q_1\left(t\right)q_2\left(t\right)}\right]\right\}+\text{const}\,.
	\end{split}
\end{equation}
Here, $z\equiv z\left(T, V\right)$ is the CE partition function of a singular source defined as 
\begin{equation}
    \label{eq:z3}
    z\left(T,V\right)=\sum_hz_h\left(T,V\right)\,,
\end{equation}
where $z_h\left(T,V\right)$ is the CE partition function of the baryon species $h$ [cf. Eq.~\eqref{eq:zh1}]. The similarity between Eq.~\eqref{eq:gB1} and Eq.~\eqref{eq:gB3} is evident. In fact, both $z_B$ and $z_{\bar{B}}$ defined in Ref. \cite{Braun-Munzinger:2020jbk} would equate to $z$ in our formalism, accounting for the resemblance of Eqs.~\eqref{eq:gB1} and \eqref{eq:gB3}. Further discussions regarding the parallelism of these two cumulant-generating functions are detailed in Appendix~\ref{AppB}.

Finally, it is worth mentioning that the derivation from Eq.~\eqref{eq:zhs2} to Eq.~\eqref{eq:gB2} has been performed without relying on specific assumptions other than the utilization of the Poisson distribution. Consequently, the result does not depend on particular thermal source details. Hence, it enables the calculation of cumulants using varied thermal sources, like the freeze-out hypersurface obtained from hydrodynamic simulations. This framework's applicability extends to scenarios involving non-conserved charge fluctuations, such as net-proton fluctuations post account for feed-down effects, provided the effects of quantum statistics are negligible~\cite{Pathria1972}. In such a case, adjustments involving the acceptance of (anti)protons become necessary, while the conditions governing global net-baryon conservation, specifically $I_B\left(\cdots\right)$ and $\sqrt{\mathcal{z}\mathcal{\bar{z}}}$ in the second line, must be preserved.
\section{Rapidity scan}\label{Sec4}

As highlighted in the Introduction, the nonmonotonic behavior in the energy dependence of high-order net-baryon cumulants stands as a promising indicator of the QCD critical point. Our previous section has revealed that these cumulants, within specific rapidity windows, are intricately tied to the acceptances of both baryons and antibaryons. This intricate relationship underscores their dependence on both the center and width of the selected rapidity window, particularly in systems manifesting inhomogeneity along the beam axis. This section primarily focuses on investigating the rapidity-dependent net-baryon cumulants in such systems. We explore their sensitivity to varying rapidity bins and assess the method of deriving effective temperature and baryon chemical potential values. This exploration aims to further advance the rapidity scan approach in probing the QCD phase diagram.
\begin{figure}[!hbtp]
    \centering
    {\includegraphics[width=0.4\textwidth]{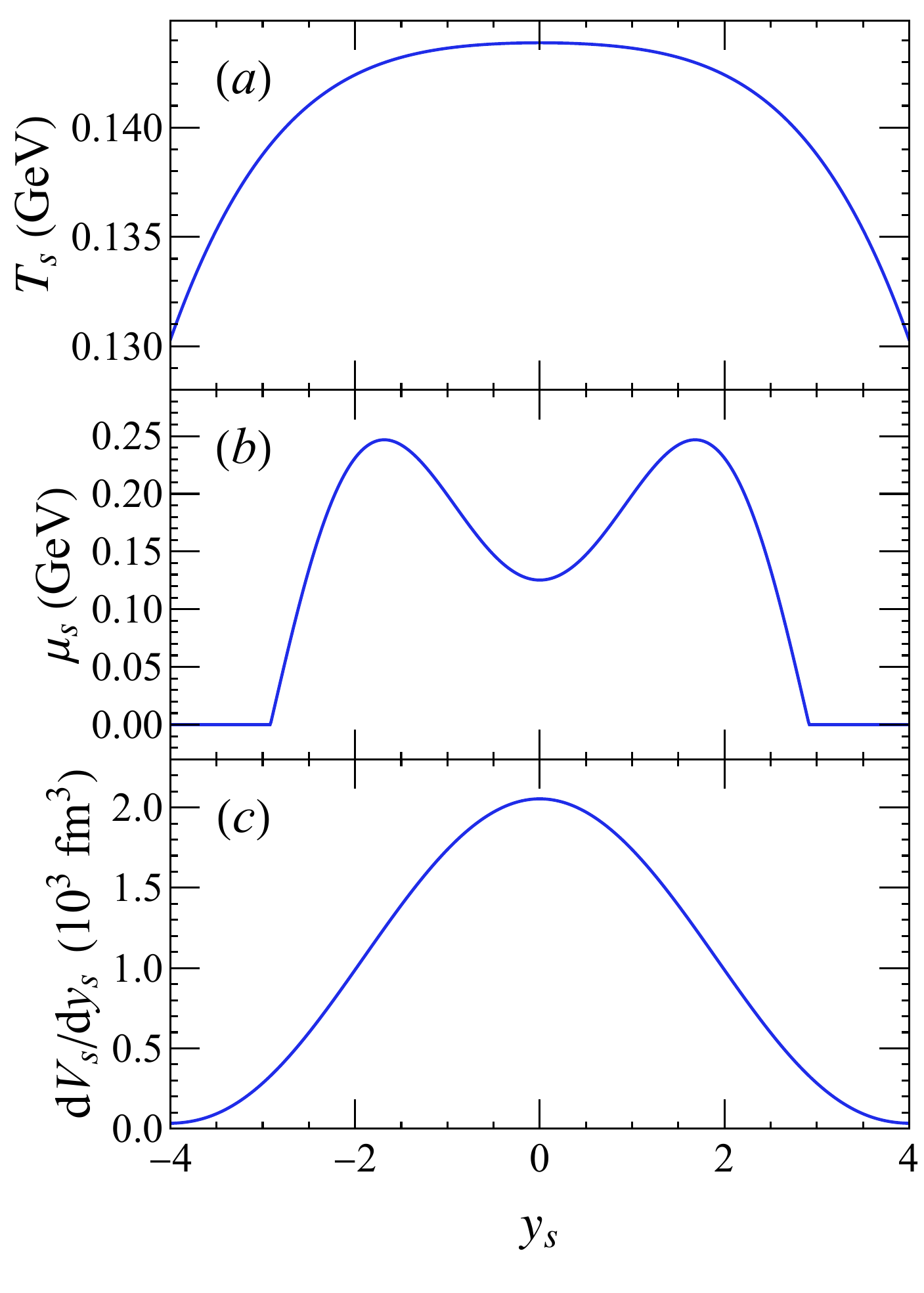}}
    {\includegraphics[width=0.4\textwidth]{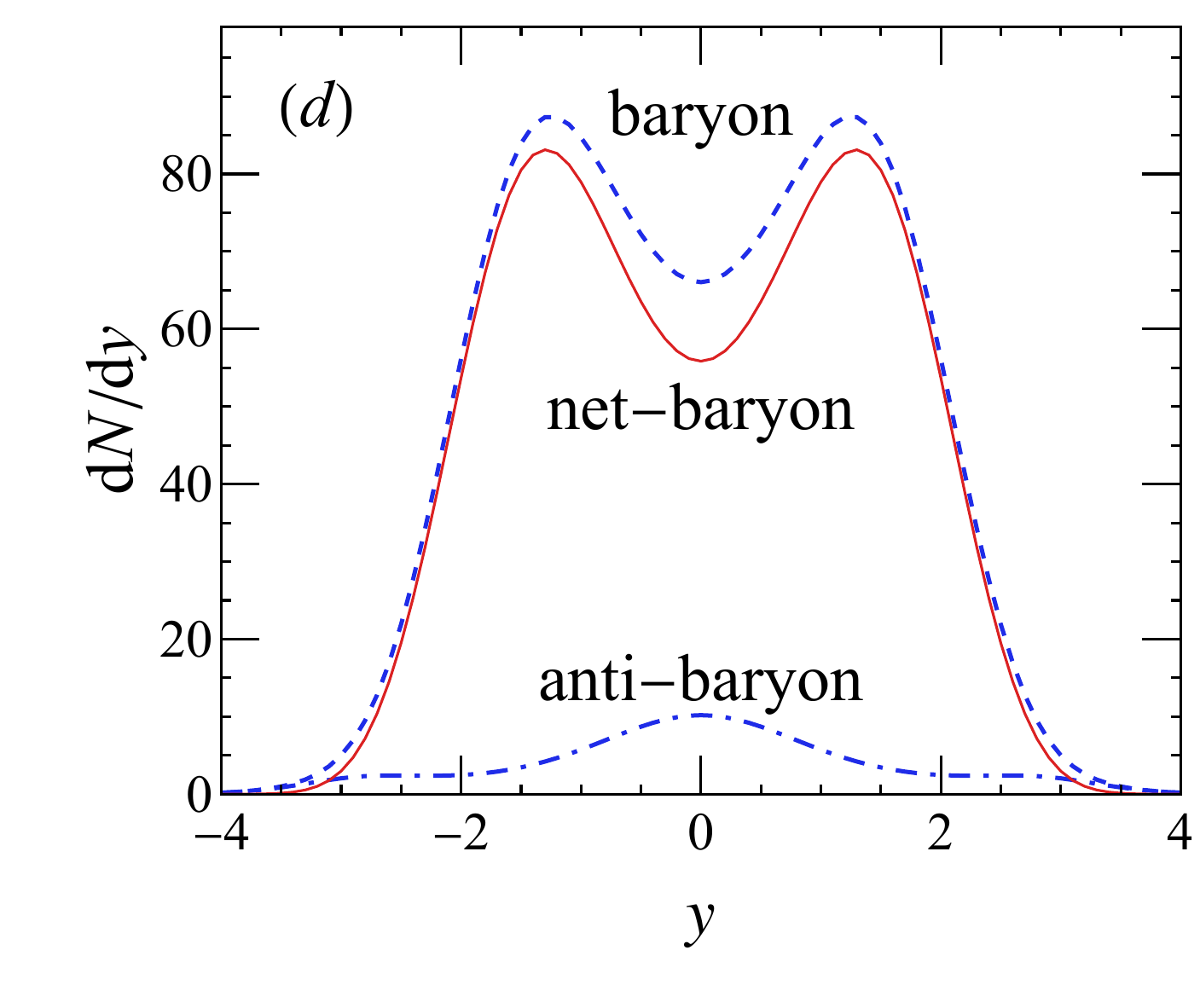}}
    \caption{Variation of thermodynamic parameters with source rapidity ($y_s$) for the thermal source at $\sqrt{s_{NN}}=19.6~\mathrm{GeV}$: ($a$) temperature ($T_s$), ($b$) baryon chemical potential ($\mu_s$), and ($c$) volume ($\dif{V_s}/\dif{y_s}$). Additionally, ($d$) illustrates the rapidity-dependent yields of baryons (blue dashed), anti-baryons (blue dot-dashed), and net-baryons (red solid).}
    \label{fig2}
\end{figure}
\subsection{Net-baryon cumulants}\label{Sec4.1}
This study focuses on investigating the rapidity scan at $\sqrt{s_{NN}}=19.6~\mathrm{GeV}$, a beam energy where various observables exhibit nonmonotonic behaviors in beam energy \cite{STAR:2017sal}. Extending the investigation to other beam energies is a natural extension, which we reserve for a future more comprehensive investigation. The thermodynamic profiles for rapidity-dependent variables associated with chemical freeze-out are displayed in FIG.~\ref{fig2}($a$)--\ref{fig2}($c$). These thermodynamic profiles are taken from FIG. 8 in Ref.~\cite{Du:2023gnv}, where they are determined by fitting the thermal yields of pions and net protons using the statistical thermal model with the consideration of thermal smearing effects.

\begin{figure}[!tbp]
    \centering
    {\includegraphics[width=0.4\textwidth]{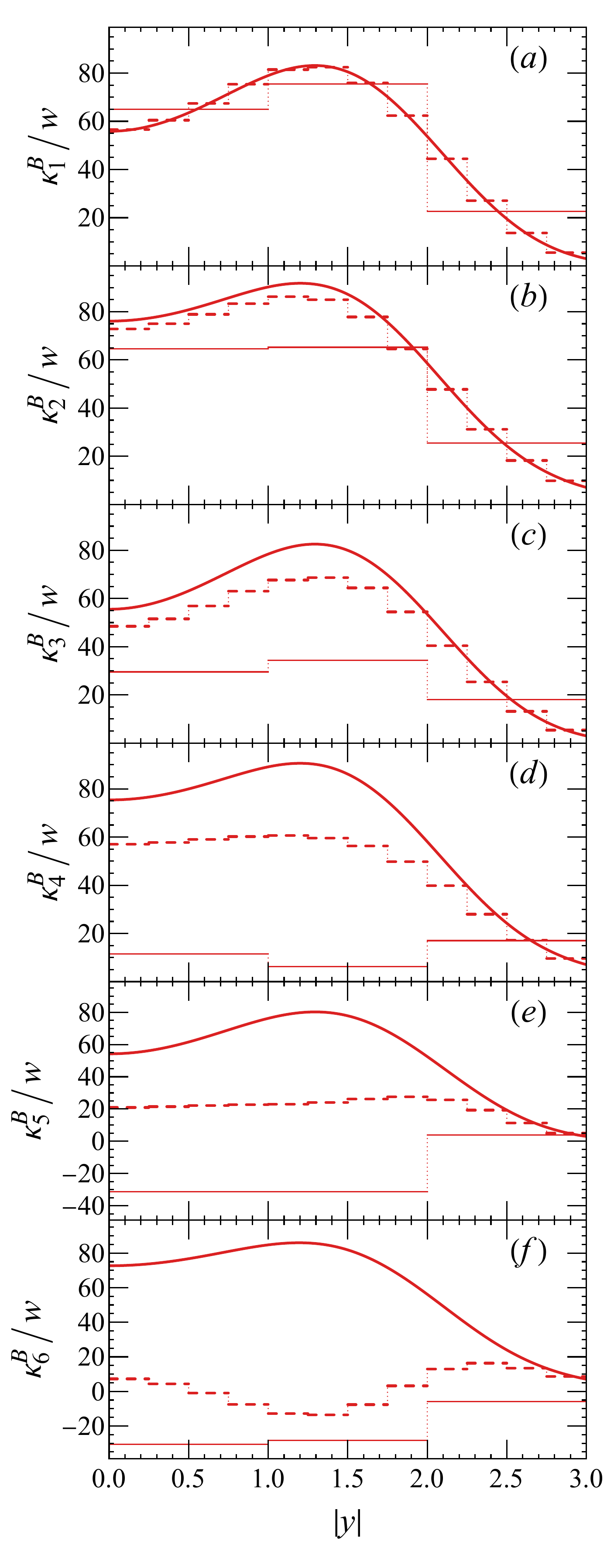}}
    \caption{Sequential representation from top to bottom of the first to sixth order scaled cumulants of net baryon fluctuation, normalized by the rapidity bin width ($w$). Each plot illustrates results corresponding to rapidity acceptance windows spanning $[0, w)$, $[w, 2w)$, and extending to $[3-w, 3]$. Three distinct bin widths are depicted: $w=0.01$ (solid lines), $w=0.25$ (dashed lines), and $w=1.00$ (thin solid lines).}
    \label{fig3}
\end{figure}

The hadron species that carry a singular baryon charge (i.e., $B_h=1$) are considered in this study (see a list summarized in Ref.~\cite{Thermal-FIST}). The resulting rapidity-dependent yields for baryons, antibaryons, and net baryons are further illustrated in FIG.~\ref{fig2} ($d$). Employing these rapidity-dependent freeze-out profiles, we investigate the behavior of cumulants by manipulating both the center and width of the rapidity acceptance window. To achieve this, we segment the $y\in[0, 3]$ range into smaller bins of equal width, forming rapidity acceptance windows such as $[0, w)$, $[w, 2w)$, and so forth up to $[3-w, 3]$. To assess the impact of bin width variations, we compare three different widths: $w=0.01$, $0.25$, and $1.00$. Figure~\ref{fig3} exhibits the first to sixth-order net-baryon cumulants normalized by the rapidity bin width ($w$). 

We observe a notable trend in FIG.~\ref{fig3}: with increasing order of cumulants, their dependence on the bin width ($w$) becomes more pronounced. This heightened sensitivity arises due to the intricate relationship between the cumulants and acceptances ($\alpha_B$ and $\alpha_{\bar B}$), as shown in Eq.~\eqref{expression of cumulant} when values for $\left<N_B\right>$ and $\left<N_{\bar{B}}\right>$ are fixed. Notably, the $\kappa_n^B$'s are $n$th-order polynomials of acceptances in Eq.~\eqref{expression of cumulant}, where the magnitudes of acceptances are roughly proportional to the width $w$. Consequently, the nonlinear terms of acceptances in $\kappa_n^B$'s become more significant in coarse rapidity scans, particularly for large bin widths such as $w=0.25$ or $1.00$. Conversely, in the fine rapidity scan ($w=0.01$), the cumulants align closely with those predicted by a Skellam distribution~\cite{Braun-Munzinger:2011shf,Braun-Munzinger:2011xux} in the case of GCE, which agrees with both mathematical and physical expectations. Mathematically, when the bin width is extremely small, a linear approximation in acceptances for cumulants becomes appropriate. As shown by Eq.~\eqref{expression of cumulant}, this approximation yields $\kappa_\text{odd}^B\approx\left<N_B\right>_A-\left<N_{\bar{B}}\right>_A$ and $\kappa_\text{even}^B \approx \left<N_B\right>_A + \left<N_{\bar{B}}\right>_A$, resulting in minimal dependence on $w$ in the case of fine scans. From a thermodynamic standpoint, the system within each rapidity bin can be considered a subsystem, and consequently, with exceedingly small bin widths, these subsystems become less sensitive to the global net-baryon number conservation governing the entire system described by CE.\footnote{%
Careful readers might have observed an oversight in our treatment regarding the rapidity bins of radiated hadrons and the associated thermal sources. However, one should note that when analyzing hadrons within a smaller bin width of $y$, their corresponding thermal sources should correspondingly stem from a smaller bin width of $y_s$.
}
This scenario parallels the conditions where the Skellam distribution becomes an appropriate description for net-baryon fluctuations, portraying baryons and antibaryons as independently fluctuating, following Poisson distributions. Further discussions on this topic can be found in Appendix~\ref{AppD}. Certainly, with an infinitesimal bin width, potential signals of critical fluctuations arising from the diverging correlation length would also vanish.

In studies of heavy-ion collisions, various models incorporating a quark-hadron crossover~\cite{Bazavov:2020bjn, Fu:2021oaw} and those excluding it~\cite{Bleicher:1999xi, Garg:2013ata} have computed net-proton cumulants. These studies found that the first- to fourth-order cumulants show consistent positivity, while the fifth- and sixth-order cumulants exhibit a negative sign in scenarios considering a quark-hadron transition. In our analysis, where the fluctuations due to quark-hadron transition are absent, we observe intriguing behaviors in $\kappa_5^B$ and $\kappa_6^B$ concerning various rapidity bin widths. As depicted in FIG.~\ref{fig3}$\left(e\right)$ and \ref{fig3}$\left(f\right)$, these cumulants maintain positivity with smaller widths and become negative as the widths increase because of the nonlinear terms that are negative in Eq.~\eqref{expression of cumulant}. Consequently, finite acceptances might influence the sign of high-order net-baryon cumulants even without considering a quark-hadron transition. We speculate that our findings on net-baryon cumulants could potentially extend to net-proton cumulants,  which are measured in experiments, at a qualitative level~\cite{Kitazawa:2011wh}. Further statistical measures inferred from low-order cumulants are detailed in FIG.~\ref{fig7} in Appendix~\ref{AppC}.

\begin{figure}[!htpb]
    \centering
    \includegraphics[width=0.44\textwidth]{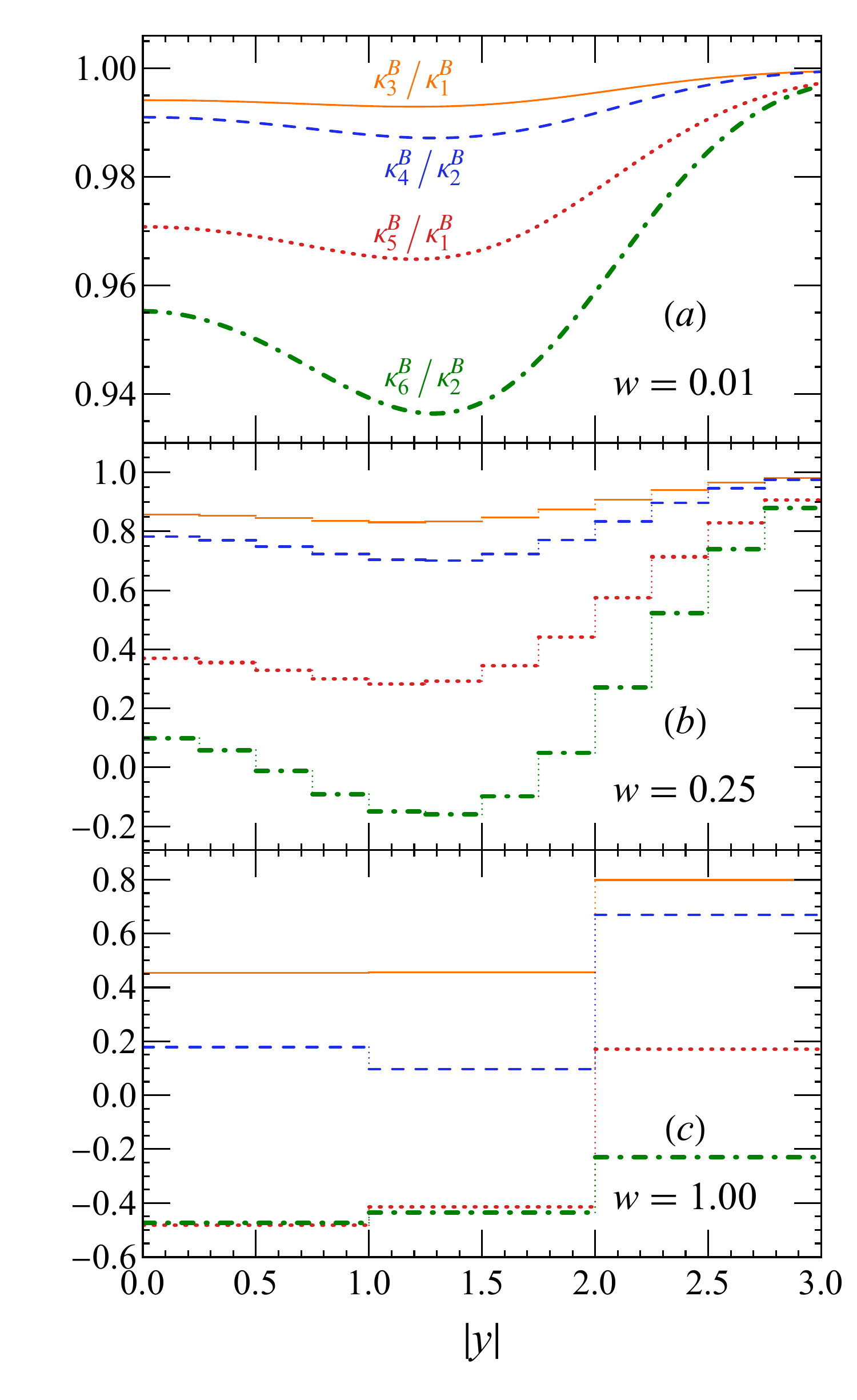}
    \caption{
    The ratios of cumulants for even and odd orders: $\kappa_3^B/\kappa_1^B$ (orange solid), $\kappa_4^B/\kappa_2^B$ (blue dashed), $\kappa_5^B/\kappa_1^B$ (red dotted), and $\kappa_6^B/\kappa_2^B$ (green dot-dashed lines). The three panels correspond to different acceptance window widths: ($a$) $w=0.01$, ($b$) $w=0.25$, and ($c$) $w=1.00$.
    }
    \label{fig4}
\end{figure}

The ratios among these cumulants, $\kappa_3^B/\kappa_1^B$, $\kappa_4^B/\kappa_2^B$, $\kappa_5^B/\kappa_1^B$, and $\kappa_6^B/\kappa_2^B$, are depicted in FIG.~\ref{fig4}. The figure distinctly illustrates their dependencies on the rapidity bin width $w$. As $w$ diminishes towards zero, these ratios converge toward their GCE limit, approaching unity, consistent with the expected behavior. An intriguing aspect lies in examining the ordering of these cumulants for different bin widths. Lattice QCD predictions indicate a specific ordering of baryon number susceptibilities $\chi_3^B/\chi_1^B>\chi_4^B/\chi_2^B>\chi_5^B/\chi_1^B>\chi_6^B/\chi_2^B$~\cite{Bazavov:2020bjn},  in contrast to ideal GCE statistical thermal model calculations that predict all ratios to be unity~\cite{Borsanyi:2018grb}. We explore this ordering in our calculations using obtained cumulants, using the fact that $\chi_m^B/\chi_n^B=\kappa_m^B/\kappa_n^B$ at a fixed temperature. In FIG.~\ref{fig4}, we observe this cascading ordering in all three cases in the rapidity window $y\in[1.0, 3.0]$ with various finite widths, indicating deviations from GCE resulting from global conservation of the net baryon number. However, in FIG. \ref{fig4}($c$), where the effects of global conservation are strong, $\kappa^B_6/\kappa^B_2$ and $\kappa^B_4/\kappa^B_2$ tend to degenerate with each other in $y\in[0, 1.0)$ and $y\in[1.0, 2.0)$. This illustrates the nontrivial dependence of the ordering on global charge conservation.

\begin{figure}[!tb]
    \centering
    {\includegraphics[width=0.4\textwidth]{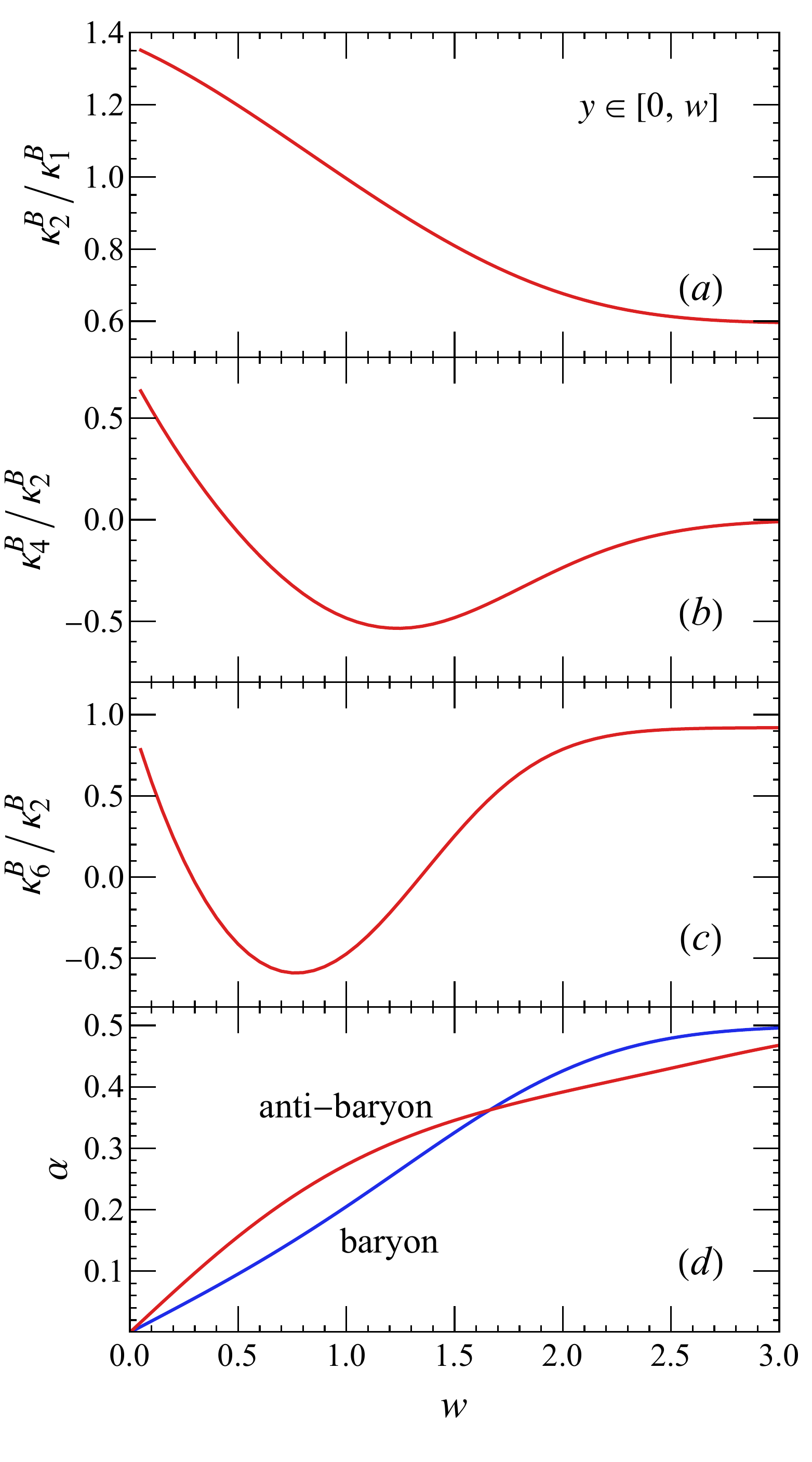}}
    \caption{Variation of the cumulant ratio  ($a$) $\kappa^B_2/\kappa^B_1$, ($b$) $\kappa^B_4/\kappa^B_2$, ($c$) $\kappa^B_6/\kappa^B_2$, and ($d$) baryon (blue) and antibaryon (red) acceptances as a function of the changing width of the rapidity window $y\in[0,w]$.}
    \label{fig5}
\end{figure}

As higher-order cumulants are more sensitive to critical points, we delve deeper into the behavior of the cumulant ratios ($\kappa^B_2/\kappa^B_1$, $\kappa^B_4/\kappa^B_2$, $\kappa^B_6/\kappa^B_2$) and explore how they vary with the changing width $w$ of the rapidity window $y\in[0,w]$. We are specifically examining the positive half of a rapidity window centered at midrapidity $y=0$ (the latter is the region where experimental data are commonly measured). Within this window, the maximum acceptance reaches 0.5 when the rapidity window coverage is sufficiently broad. Figure~\ref{fig5} illustrates that as the rapidity window widens and thus substantial acceptance coverage is approached, the $\kappa_6^B/\kappa_2^B$ ratio reaches a plateau, saturating towards a value near unity. This phenomenon occurs as the acceptances approach 0.5, indicating the inclusion of most particles and antiparticles. Intriguingly, the behavior of $\kappa_6^B/\kappa_2^B$ is nonmonotonic and changes sign when the rapidity window has $w \in \left[0.5,1.0\right]$, solely due to the increasing bin width and corresponding rising acceptances, despite the absence of quark-hadron transition considered in our calculation. Figure~\ref{fig5}, together with FIGs.~\ref{fig3} and \ref{fig4}, illustrates the significant impact of finite acceptance on the noncritical yet nontrivial behaviors observed in net-baryon cumulants. Some of these observed behaviors exhibit characteristics resembling nonmonotonic trends typically associated with critical effects. Therefore, these results highlight the crucial importance of accounting for finite acceptance, when investigating the high-order net-proton cumulants within a limited rapidity window.

From the perspective of the beam energy scan method, high-order net-proton cumulants are commonly measured within a finite rapidity window centered around midrapidity to study the fluctuations at a length scale comparable to the correlation length, spanning different beam energies. For instance, the STAR Collaboration measures net-proton cumulants within the rapidity range $|y|{\,<\,}1$. However, at lower beam energies characterized by smaller beam rapidities, the acceptances of particles are expected to increase within a fixed rapidity coverage when the beam energy decreases. As seen in FIG.~\ref{fig5}, the varying acceptances across different beam energies may impact the observed behaviors of high-order cumulants. Therefore, careful consideration is necessary when attempting to discern potential critical phenomena from the nonmonotonic behaviors observed in net-proton cumulants at low beam energies or from their deviation from the noncritical statistical thermal baseline~\cite{STAR:2020tga, STAR:2021iop}.
\subsection{Extraction of thermodynamic variables}\label{Sec4.2}
Having analyzed the rapidity-dependent net-baryon cumulants and their intricate relationship with varying rapidity bin widths, we turn to another pertinent exploration: extracting the thermodynamic characteristics of a system exhibiting inhomogeneity along the beam axis from net-baryon cumulants obtained within distinct rapidity windows, such as those from experimental measurements. Ideally, the derived effective thermodynamic properties should accurately replicate the given (or measured) net-baryon cumulants.\footnote{%
Of course, it is possible to derive the effective thermodynamic properties by matching the identified hadron yields. However, in this study, our primary objective is to reproduce the net-baryon cumulants.
} 
To address this issue, we will leverage the relationship we discussed between Eqs.~\eqref{eq:gB1} and~\eqref{eq:gB3}. The former pertains to the generating function within a particular rapidity window, while the latter describes a singular source with specific temperature and baryon chemical potential. If we identify a singular thermal source capable of producing the same generating function as that within a rapidity window, it can effectively reproduce the corresponding net-baryon cumulants.

\begin{figure}[!bpt]
    {\includegraphics[width=0.4\textwidth]{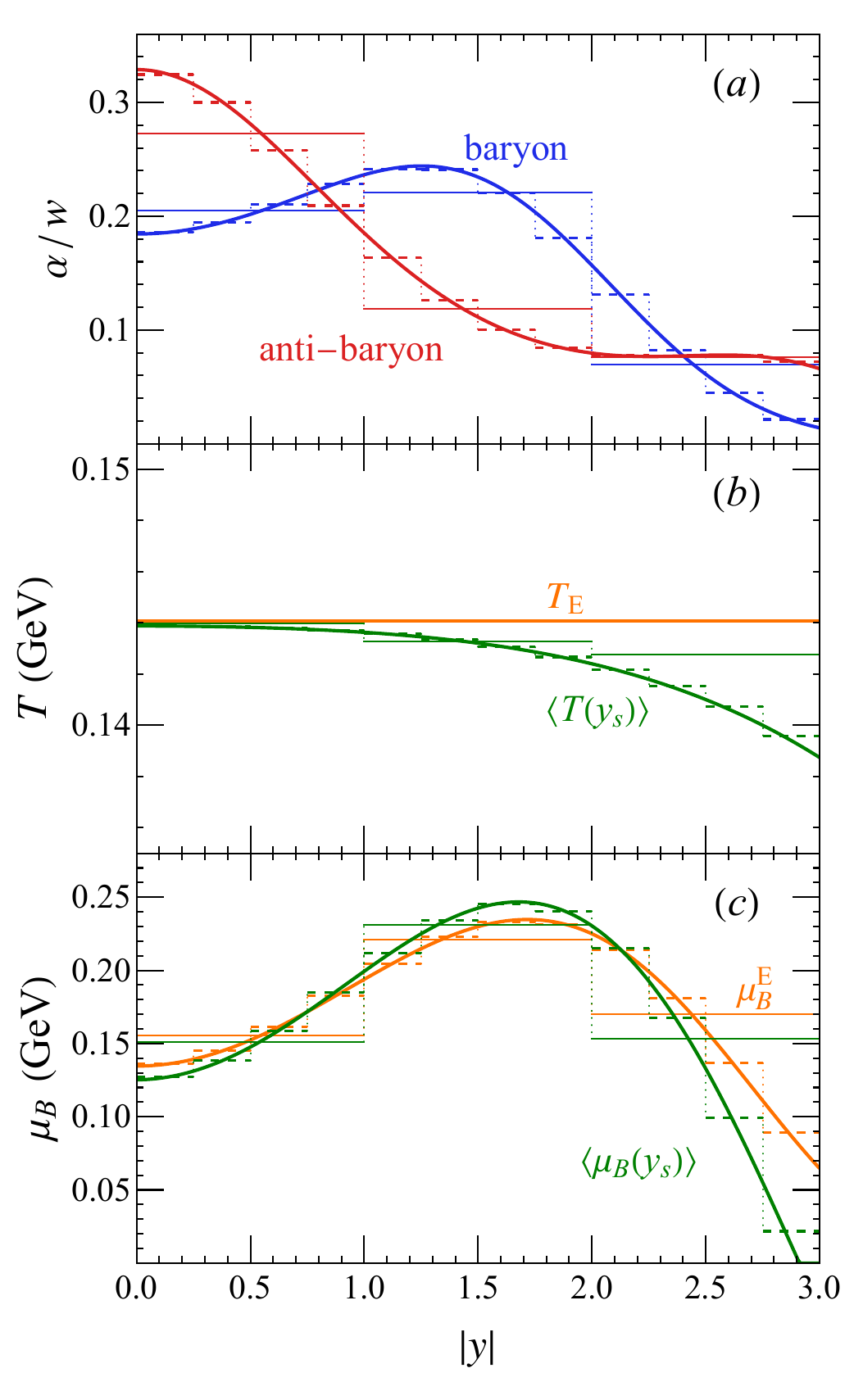}}
    \caption{($a$) Scaled acceptances of baryons (blue) and antibaryons (red), ($b$) effective $T_\mathrm{E}$ (orange) and average $\left<T\left(y_s\right)\right>$ (green) temperatures, as well as ($c$) effective $\mu_B^\mathrm{E}$ (orange) and average  $\left<\mu_B\left(y_s\right)\right>$ (green) baryon chemical potentials. Their variations across three different rapidity bins are also illustrated: $w{\,=\,}0.01$ (solid), $w{\,=\,}0.25$ (dashed), and $w{\,=\,}1.00$ (thin solid lines).}
    \label{fig6}
\end{figure}

Specifically, considering the net-baryon cumulants within a rapidity window of $[y_c-w/2, y_c+w/2]$, we position a thermal source at $y_s{\,=\,}y_c$, aligning with the center of the target window. This thermal source, characterized by its temperature $(T_\mathrm{E})$, baryon chemical potential $(\mu_B^\mathrm{E})$, and volume $(V_\mathrm{E})$, with the superscript or subscript $\mathrm{E}$ denoting their representation as effective thermodynamic variables of the rapidity window, has the rapidity-differential single-particle partition function outlined in Eq.~\eqref{eq:zhs1} for a single source. A comparison between Eqs.~\eqref{eq:gB1} and~\eqref{eq:gB3} reveals that for the generating function of the singular thermal source to match that of the rapidity window,  the effective temperature $(T_\mathrm{E})$, effective baryon chemical potential $(\mu_B^\mathrm{E})$, and effective volume $(V_\mathrm{E})$ should be adjusted so that the former can replicate the acceptances of both baryons and antibaryons observed in the latter:
\begin{equation}
\label{alpha_E}
\alpha_B^\mathrm{E}\left(\mu_B^\mathrm{E},T_\mathrm{E},V_\mathrm{E}\right)=\alpha_B\,,\quad	\alpha_{\bar{B}}^\mathrm{E}\left(\mu_B^\mathrm{E},T_\mathrm{E},V_\mathrm{E}\right)=\alpha_{\bar{B}}\,.
\end{equation}
Results depicting the acceptances [$\alpha_B$ and $\alpha_{\bar{B}}$ on the right-hand side of Eq.~\eqref{alpha_E}] within various rapidity bins for three distinct bin widths are presented in FIG.~\ref{fig6}($a$). Indeed, as anticipated, the distributions of baryon and anti-baryon acceptances follow the distributions of their respective yields as shown in FIG. \ref{fig2}($d$).
Besides, the partition functions should satisfy
\begin{equation}
\label{Eq1}
	z\left(T_\mathrm{E},\,V_\mathrm{E}\right)=\sqrt{\mathcal{z}\bar{\mathcal{z}}}\,,
\end{equation}
where on the left-hand side the partition function is given by Eq.~\eqref{eq:z3} for a singular source, while the right-hand side a term from Eq.~\eqref{eq:gB1}, is a quantity defined for the entire system exhibiting inhomogeneity along the beam axis. 

Ideally, to obtain the effective thermodynamic parameters within each rapidity bin, it is necessary to fine-tune the parameters such that they satisfy both Eqs.~\eqref{alpha_E} and \eqref{Eq1}. However, as evident from FIG.~\ref{fig2}($a$), the temperature variation along the rapidity axis at $\sqrt{s_{NN}} = 19.6$~GeV is notably small. Given this observation, we will aim to simplify the process by estimating a constant effective temperature across all rapidity bins.\footnote{%
It should be noted that this simplification might not be suitable at lower beam energies characterized by more pronounced temperature variations.
} 
To achieve this, we position a singular thermal source at $y_s=0$ to represent the entire fireball. We require that the effective temperature $T_\mathrm{E}$ and volume $V_\mathrm{E}^\mathrm{tot}$ of this source have specific values that satisfy Eq.~\eqref{Eq1}. Here, we employ the total volume of the entire fireball, as depicted in FIG.~\ref{fig2}($c$), to estimate the effective volume value, namely, $V_\mathrm{E}^\mathrm{tot}=\int_{-\infty}^{+\infty}\dif y_s\, \dif V_s/\dif y_s$. By substituting $V_\mathrm{E}^\mathrm{tot}$ into Eq.~\eqref{Eq1}, we derive the effective temperature $T_\mathrm{E}\approx144.1$~MeV for $\sqrt{s_{NN}}=19.6$~GeV. Admittedly, our method of deriving the effective volume relies on information that is unknown from the experimental measurements and thus should be improved in the future. However, this oversight is acceptable as our primary focus lies in extracting the effective chemical potential exhibiting more pronounced variation from the net-baryon cumulants. 

Now, employing the acquired $T_\mathrm{E}\approx144.1$~MeV as the effective temperature across all rapidity bins [see FIG.~\ref{fig6}($b$)], our next step involves extracting the effective baryon chemical potential using Eqs.~\eqref{alpha_E}. The acceptances of baryons and anti-baryons corresponding to the singular thermal source representing a specific rapidity bin $[y_c-w/2,y_c+w/2]$ are respectively determined by
\begin{equation}   
	\begin{gathered}  
		\alpha_B^\mathrm{E}\left(\mu_B^\mathrm{E},T_\mathrm{E},V_\mathrm{E}\right)=\frac{\left<N_B^\mathrm{E}\right>_A}{\left<N_B\right>}\,,\\	\alpha_{\bar{B}}^\mathrm{E}\left(\mu_B^\mathrm{E},T_\mathrm{E},V_\mathrm{E}\right)=\frac{\left<N_{\bar{B}}^\mathrm{E}\right>_A}{\left<N_{\bar{B}}\right>}\,.
	\end{gathered}
\end{equation}
Within the rapidity window, the numbers of baryons and anti-baryons can be computed respectively using
\begin{equation}\label{NBE}
	\begin{gathered}
		\left<N_B^\mathrm{E}\right>_A=\sum_h\int_{y_c-\frac{w}{2}}^{y_c+\frac{w}{2}}\e^{\frac{\mu_B^\mathrm{E}}{T_\mathrm{E}}}z_h\left(T_\mathrm{E},V_\mathrm{E};y-y_c\right)\dif y\,,\\
		\left<N_{\bar{B}}^\mathrm{E}\right>_A=\sum_h\int_{y_c-\frac{w}{2}}^{y_c+\frac{w}{2}}\e^{-\frac{\mu_B^\mathrm{E}}{T_\mathrm{E}}}z_h\left(T_\mathrm{E},V_\mathrm{E};y-y_c\right)\dif y\,,
	\end{gathered}
\end{equation}
where $z_h$ is the partition function of a hadron species introduced in Eq.~\eqref{eq:zhs1}. In principle, the effective volume in Eq.~\eqref{NBE} could also be adjusted individually for each rapidity bin, but our primary focus lies in determining the baryon chemical potential, hence we refrain from explicitly extracting the volume. For this purpose, replicating the ratio of baryon and anti-baryon acceptances within the rapidity window in the singular source is sufficient, i.e.,
\begin{equation}
	\label{Eq2}
\frac{\alpha_B^\mathrm{E}\left(\mu_B^\mathrm{E},T_\mathrm{E},V_\mathrm{E}\right)}{\alpha_{\bar{B}}^\mathrm{E}\left(\mu_B^\mathrm{E},T_\mathrm{E},V_\mathrm{E}\right)}=\frac{\alpha_B}{\alpha_{\bar{B}}}\,,
\end{equation}
where volume is canceled out when calculating the ratio. The effective chemical potential can be derived by solving Eq.~\eqref{Eq2}, and the results for three distinct bin widths  are presented in FIG.~\ref{fig6}($c$).

To evaluate the effective thermodynamic variables derived from the rapidity-dependent net-baryon cumulants, we can compare them against the thermodynamic profiles depicted in FIG.~\ref{fig2}, which are used for obtaining these cumulants. Given the varying nature of these profiles across rapidity, we shall define the average thermodynamic variables, including the average temperature $\left<T\left(y_s\right)\right>$ and baryon chemical potential $\left<\mu_B(y_s)\right>$, based on the smooth profiles within distinct rapidity bins. For a rapidity window $y_s\in\left[y_c-w/2,y_c+w/2\right]$, we calculate these average variables by requiring them to yield identical GCE partition functions and net-baryon values as calculated from the smooth profiles:
\begin{equation}
	\begin{split}
	&2\cosh\left(\frac{\left<\mu_B(y_s)\right>}{\left<T\left(y_s\right)\right>}\right)z\left(\left<T\left(y_s\right)\right>,\int_{y_c-\frac{w}{2}}^{y_c+\frac{w}{2}}\frac{\dif{V}}{\dif{y_s}}\dif{y_s}\right)
        \\=\;& \int_{y_c-\frac{w}{2}}^{y_c+\frac{w}{2}}\dif{y_s}\int_{-\infty}^{+\infty}\dif{y}\sum_h\left(\mathcal{z}_h^s+\bar{\mathcal{z}}_h^s\right)\,,\\
	&2\sinh\left(\frac{\left<\mu_B\left(y_s\right)\right>}{\left<T\left(y_s\right)\right>}\right)z\left(\left<T\left(y_s\right)\right>,\int_{y_c-\frac{w}{2}}^{y_c+\frac{w}{2}}\frac{\dif{V}}{\dif{y_s}}\dif{y_s}\right)
        \\=\;&\int_{y_c-\frac{w}{2}}^{y_c+\frac{w}{2}}\dif{y_s}\int_{-\infty}^{+\infty}\dif{y}\sum_h\left(\mathcal{z}_h^s-\bar{\mathcal{z}}_h^s\right)\,.
	\end{split}
\end{equation}
Here, the left-hand side denotes the GCE partition function and the resulting net-baryon number calculated for a singular source, whereas the right-hand side corresponds to the corresponding values derived from the continuous freeze-out profiles presented in FIG.~\ref{fig2}. The average variables across distinct rapidity windows for three different rapidity bin widths are illustrated in FIG.~\ref{fig6}. These average variables converge toward the continuous freeze-out profiles depicted in FIG.~\ref{fig2} when the bin widths become exceedingly small. 

FIG.~\ref{fig6} ($b$) illustrates an intriguing trend that emerges with increasing bin width $w$: the average temperature $\left<T\left(y_s\right)\right>$ within a rapidity window appears to approach the highest temperature observed within the same bin. It also shows that the constant effective temperature $T_\mathrm{E}$ approximately matches the highest temperature within the continuous freeze-out temperature profile. These trends can be understood by accounting for the influence of total net-baryon conservation across the entire system, as outlined in Eq.~\eqref{eq:zhs2}. Within a rapidity bin governed by the GCE, the fluctuations of baryons and anti-baryons contribute to the free energy of the (sub)system, represented by the fugacity factor $\exp\left(\mu_B/T\right)$. In contrast, the CE suppresses these fluctuations, a scenario that becomes more pronounced with larger rapidity bins. In our method of extraction, achieving equivalent thermodynamic properties requires compensating for the energy contributed by fluctuations absent in CE. Therefore, a higher temperature is necessary to align these properties with larger rapidity bins. For a deeper understanding of this concept, interested readers may refer to FIG. 1 in Ref. \cite{Vovchenko:2020tsr}. FIG.~\ref{fig6}($c$) demonstrates closely matching values between the derived chemical potential ($\mu_B^\mathrm{E}$) and the averaged value ($\left<\mu_B\left(y_s\right)\right>$) across distinct rapidity bins, showcasing small differences partially attributed to the thermal smearing effect. Moreover, the rapidity scan can capture variations in the thermodynamic properties of the fireball along the beam direction. Figure~\ref{fig6} illustrates a close agreement between the effective thermodynamic variables obtained from the rapidity-dependent net-baryon cumulants and the averaged values derived from the continuous profiles showcased in FIG.~\ref{fig2}, which underlines the efficacy of the extraction method. However, it is worth noting that due to critical slowing down, considering the off-equilibrium evolution of fluctuations in rapidity scans becomes necessary \cite{Nahrgang:2018afz,Nahrgang:2020yxm,Mukherjee:2015swa, Mukherjee:2016kyu, Bluhm:2018qkf, Bluhm:2019yfb, Kitazawa:2020kvc,Rajagopal:2019xwg,Du:2020bxp,  Pradeep:2021opj, Sogabe:2021svv, Hammelmann:2022yso, Pradeep:2022mkf, Pradeep:2022eil, Hammelmann:2023aza}. This aspect is essential for the search for the critical point in heavy-ion collisions.
\section{Summary and outlook}\label{Sec5}

The upcoming BES-II measurements with improved statistical precision and expanded rapidity-dependent data, will play a crucial role in searching for the critical phenomena, especially through high-order net-proton cumulants across a broad range of beam energies. In this study, we explored the statistical thermal model which integrates rapidity-dependent freeze-out profiles to characterize variances in thermodynamic properties along the beam axis. 

We derived analytical expressions for net-baryon cumulants within finite rapidity ranges, maintaining total net-baryon conservation under the canonical ensemble for the entire system. The exploration of rapidity-dependent net-baryon cumulants at $\sqrt{s_{NN}}=19.6$ GeV underscored the substantial influence of finite acceptance on their behavior with varying rapidity windows. As depicted in Figs.~\ref{fig3}, \ref{fig4}, and \ref{fig5}, these kinematic acceptances significantly affected high-order cumulant behaviors, unveiling nontrivial trends which can be similar to characteristics associated with critical effects. This observation strongly advocates for considering finite acceptance when interpreting high-order net-proton cumulants measured within confined rapidity windows. 

In the context of the Beam Energy Scan, where these cumulants are observed within a fixed rapidity range around midrapidity, variations in particle acceptances due to varying beam rapidity across different beam energies are expected to emerge as pivotal factors. Therefore, careful investigation is essential when searching for potential critical phenomena through the energy dependence of net-proton cumulants. Additionally, our study investigated methods to derive effective temperature and baryon chemical potential values, advancing the rapidity scan approach in probing the QCD phase diagram.

Future studies will refine extraction methods for thermodynamic variables, integrating identified hadrons with feed-down effects. Incorporating both critical phenomena (quark-hadron transition, non-Gaussian fluctuations) and noncritical contributions (nonequilibrium corrections, volume fluctuations) will further improve the rapidity scan utilizing Beam Energy Scan measurements and facilitate unraveling phenomena associated with the QCD critical point.
\section*{Acknowledgment}

We are grateful for the helpful discussion with Drs.\, Heng-Tong Ding, Xiaofeng Luo, Jianqiao Wang, and Yapeng Zhang. This work is supported in part by the Institute of Modern Physics of the Chinese Academy of Sciences under grant No. E11S641GR0 (J.~L.), in part by the Natural Sciences and Engineering Research Council of Canada (L.~D.), and in part by Tsinghua University under grant No. 53330500923 (S.~S.).

\appendix
\section{Single-particle partition functions table}\label{AppA}

Table~\ref{Conventions} provides a comprehensive summary detailing the symbols and interpretations of various partition functions.
\begin{table}[!htbp]
\begin{tabular}{ccc}
        \hline
		\hline
		Convention & Eq.  & Single-particle partition function of\\
		\hline
		$z_h^s$  & \eqref{eq:zh1}
		& 
		$h$ radiated from $s$ in CE \\
		$z_h^s\left(y\right)$ & \eqref{eq:zhs1}
		&
            \thead{$h$ radiated from $s$ \\ distributing on $y$ in CE}\\
		$\mathcal{z}_h^s\left(y-y_s\right)$ & \eqref{eq:zhs2} 
		&
		\thead{baryon $h$ at $y$ radiated\\from $s$ at $y_s$ in GCE} \\	
		$\bar{\mathcal{z}}_h^s\left(y-y_s\right)$ & \eqref{eq:zhs2}
		&
		\thead{antibaryon $h$ at $y$ radiated\\from $s$ at $y_s$ in GCE}\\
		$\mathcal{z}_{A,h}^s$  & \eqref{eq:zAhs1}
		&
		\thead{baryon $h$ in acceptance\\radiated from $s$ at $y_s$ in GCE} \\	
		$\bar{\mathcal{z}}_{A,h}^s$ & \eqref{eq:zAhs1}
		&
		\thead{antibaryon $h$ in acceptance\\radiated from $s$ at $y_s$ in GCE} \\
		$\mathcal{z}_{R,h}^s$ & \eqref{eq:zRhs1}
		&
		\thead{baryon $h$ outside acceptance\\radiated from $s$ at $y_s$ in GCE} \\	
		$\bar{\mathcal{z}}_{R,h}^s$ & \eqref{eq:zRhs1}
		&
		\thead{antibaryon $h$ outside acceptance\\radiated from $s$ at $y_s$ in GCE} \\
		$\mathcal{z}_A$ & \eqref{eq:z2}
		&
		\thead{baryons in acceptance in GCE} \\	
		$\bar{\mathcal{z}}_A$ & \eqref{eq:z2}
		&
		\thead{antibaryons in acceptance in GCE} \\
		$\mathcal{z}_R$ & \eqref{eq:z2}
		&
		\thead{baryons outside acceptance in GCE} \\	
		$\bar{\mathcal{z}}_R$ & \eqref{eq:z2}
		&
		\thead{antibaryons outside acceptance in GCE} \\
		$\mathcal{z}$ & \eqref{eq:gB2}
		&
		\thead{baryons in full phase space in GCE} \\	
		$\bar{\mathcal{z}}$ & \eqref{eq:gB2}
		&
		\thead{antibaryons in full phase space in GCE} \\
        $z_h$ & \eqref{eq:z3}
        &
        $h$ in single-source model in CE\\
        $z$  & \eqref{eq:z3}
        &
        single-source model in CE\\
		\hline
        \hline
\end{tabular}
    \caption{Conventions of single-particle partition functions.\label{Conventions}}
\end{table}
\section{On the connection between Eq.~\eqref{eq:gB1} and Eq.~\eqref{eq:gB3}} \label{AppB}

As previously mentioned, the variables ${\mathcal{z}}$, $\bar{\mathcal{z}}$, $\alpha_{{B}}$, and $\alpha_{\bar{B}}$ in Eq.~\eqref{eq:gB1} pertain to a system characterized by rapidity inhomogeneity. Conversely, the quantities outlined in Eq.~\eqref{eq:gB3} are computed for a homogeneous or singular thermal source. To bridge the connection between these two generating functions, we will introduce a few additional approximations.

In CE, the expected values of baryon and anti-baryon numbers for a singular source are determined by~\cite{Braun-Munzinger:2020jbk}
\begin{equation}
    \left<N_B\right>=z\frac{I_{B-1}\left(2z\right)}{I_{B}\left(2z\right)}\,,\qquad 
    \left<N_{\bar{B}}\right>=z\frac{I_{B+1}\left(2z\right)}{I_{B}\left(2z\right)}\,.
\end{equation}
In the high-energy limit, where $\left<N_B\right>\approx\left<N_{\bar{B}}\right>\gg B$, the total net-baryon number $B\approx0$. Utilizing the asymptotic form of $I_\nu\left(x\right)$ for $x\gg\left|\nu^2-1/4\right|$~\cite{Braun-Munzinger:2020jbk},
\begin{equation}
    \begin{split}
        \lim_{x\gg\left|\nu^2-\frac{1}{4}\right|}I_\nu\left(x\right)&=\frac{\e^{x}}{\sqrt{2\pi x}}\left(1-\frac{4\nu^2-1}{8x}
        \cdots\right)\,,
    \end{split}
\end{equation}
we obtain $\left<N_B\right>\approx \left<N_{\bar{B}}\right>\approx z$. In low-energy collisions, utilizing the asymptotic form of $I_\nu\left(z\right)$ at $\nu\rightarrow\infty$~\cite{Olver2010nist},
\begin{equation}
    \label{low energy approximation}
    \lim_{\nu\rightarrow\infty}I_\nu\left(x\right)=\frac{1}{\sqrt{2\pi\nu}}\left(\frac{\e x}{2\nu}\right)^\nu,\quad x\neq0\,,
\end{equation}
we obtain $\left<N_B\right>\approx B$ and $\left<N_{\bar{B}}\right>\approx z^2/B$ because $B\gg1$. In both scenarios, we can consistently observe that $z\approx\sqrt{\left<N_B\right>\left<N_{\bar{B}}\right>}$ for a singular thermal source within CE. 

In the main text, we have shown that when solely considering baryons carrying a single baryon charge and excluding light nuclei with $B_h>1$, $\mathcal{z}$ and $\bar{\mathcal{z}}$ yield the expected values of baryons $\left<N_B\right>$ and anti-baryons $\left<N_{\bar{B}}\right>$ in the full phase space. Consequently, in a system characterized by rapidity inhomogeneity, it follows that $\sqrt{\mathcal{z}\bar{\mathcal{z}}}\approx\sqrt{\left<N_B\right>\left<N_{\bar{B}}\right>}$. Thus, it can be seen that both $\sqrt{\mathcal{z}\bar{\mathcal{z}}}\approx\sqrt{\left<N_B\right>\left<N_{\bar{B}}\right>}$ in Eq.~\eqref{eq:gB1} for a system characterized by rapidity inhomogeneity and $z\approx\sqrt{\left<N_B\right>\left<N_{\bar{B}}\right>}$ in Eq.~\eqref{eq:gB3} for a singular source reflect the characteristics of the CE.
\section{Expressions of the first to sixth order cumulants and statistical measures}\label{AppC}

For clarity and coherence within this study, we present the analytical expressions of cumulants up to the sixth order in this section. It is worth noting that this formalism is introduced and developed in Ref. \cite{Braun-Munzinger:2020jbk}.
With the notations
\begin{align}
        \kappa_1^\pm&=
        \frac{1}{2}\left(\alpha_B\pm\alpha_{\bar{B}}\right)\,,\notag\\ 
	\kappa_2^\pm&=
        \frac{1}{2}\left[\alpha_B\left(1-\alpha_B\right)\mp\left(\alpha_B\leftrightarrow\alpha_{\bar{B}}\right)\right]\,,\notag\\
	\kappa_3^\pm&=
        \frac{1}{2}\left[\alpha_B\left(1-\alpha_B\right)\left(1-2\alpha_B\right)\pm\left(\alpha_B\leftrightarrow\alpha_{\bar{B}}\right)\right]\,,\notag\\
	\kappa_4^\pm&=
        \frac{1}{2}\left[\alpha_B\left(1-\alpha_B\right)\left(1-6\alpha_B+6\alpha_B^2\right)\mp\left(\alpha_B\leftrightarrow\alpha_{\bar{B}}\right)\right]\,,\\
	\kappa_5^\pm&=
        \frac{1}{2}\,\Big[\alpha_B\left(1-\alpha_B\right)\left(1-2\alpha_B\right)\left(1-12\alpha_B+12\alpha_B^2\right)\notag\\
        &\qquad
        \pm\left(\alpha_B\leftrightarrow \alpha_{\bar{B}}\right)\Big]\,,\notag\\
	\kappa_6^\pm&=
        \frac{1}{2}\,\Big[\alpha_B\left(1-\alpha_B\right)\left(1-30\alpha_B+150\alpha_B^2-240\alpha_B^3+120\alpha_B^4\right)
        \notag\\
        &\qquad
        \mp\left(\alpha_B\leftrightarrow \alpha_{\bar{B}}\right)\Big]\,,\notag
\end{align}
the first to sixth-order cumulants can be expressed as
\begin{equation}
	\label{expression of cumulant}
	\begin{split}
	    \kappa_1^B&=
        B\kappa_1^++S\kappa_1^-\,,\\
	\kappa_2^B&=
        B\kappa_2^++S\kappa_2^-\,,\\
	\kappa_3^B&=
        B\kappa_3^++S\kappa_3^-+8P\left(\kappa_1^-\right)^3\,,\\
	\kappa_4^B&=
        B\kappa_4^++S\kappa_4^-+48P\left(\kappa_1^-\right)^2\kappa_2^--16P\left(S-1\right)\left(\kappa_1^-\right)^4\,,\\
	\kappa_5^B&=
        B\kappa_5^++S\kappa_5^-+40P\left[3\kappa_1^-\left(\kappa_2^-\right)^2+2\left(\kappa_1^-\right)^2\kappa_3^-\right]
        \\&\quad
        -160P\left(S-1\right)\left(\kappa_1^-\right)^3\kappa_2^-+32P\left(S^2-S+1\right)\,,\\
	\kappa_6^B&=
        B\kappa_6^++S\kappa_6^-+120P\left[4\kappa_1^-\kappa_2^-\kappa_3^-+\left(\kappa_1^-\right)^2\kappa_4^-+\left(\kappa_2^-\right)^3\right]
        \\&\quad
        -80P\left(S-1\right)\left[4\left(\kappa_1^-\right)^3\kappa_3^-+9\left(\kappa_1^-\right)^2\left(\kappa_2^-\right)^2\right]
        \\&\quad
        +480P\left(S^2-S+1\right)\left(\kappa_1^-\right)^4\kappa_2^-
        \\&\quad
        -64P\left[\left(S^3-S^2+S-1\right)+6P\right]\left(\kappa_1^-\right)^6\,.
	\end{split}
\end{equation}
Here we have employed the same abbreviations introduced in Ref.~\cite{Braun-Munzinger:2020jbk}, summarized below:
\begin{equation}
    S=\left<N_B\right>+\left<N_{\bar{B}}\right>\,,\qquad
    P=\left<N_B\right>\left<N_{\bar{B}}\right>\,.
\end{equation}
Some abbreviations may yield different results compared to those shown in Ref. \cite{Braun-Munzinger:2020jbk}. This discrepancy arises due to the condition $\mathcal{z}\bar{\mathcal{z}}\approx\left<N_B\right>\left<N_{\bar{B}}\right>$ within our framework, resulting in $Q=\mathcal{z}\bar{\mathcal{z}}-P\approx0$ and $W=QS-P\approx-P$.
\begin{figure}[!htp]
	\centering
	{\includegraphics[width=0.4\textwidth]{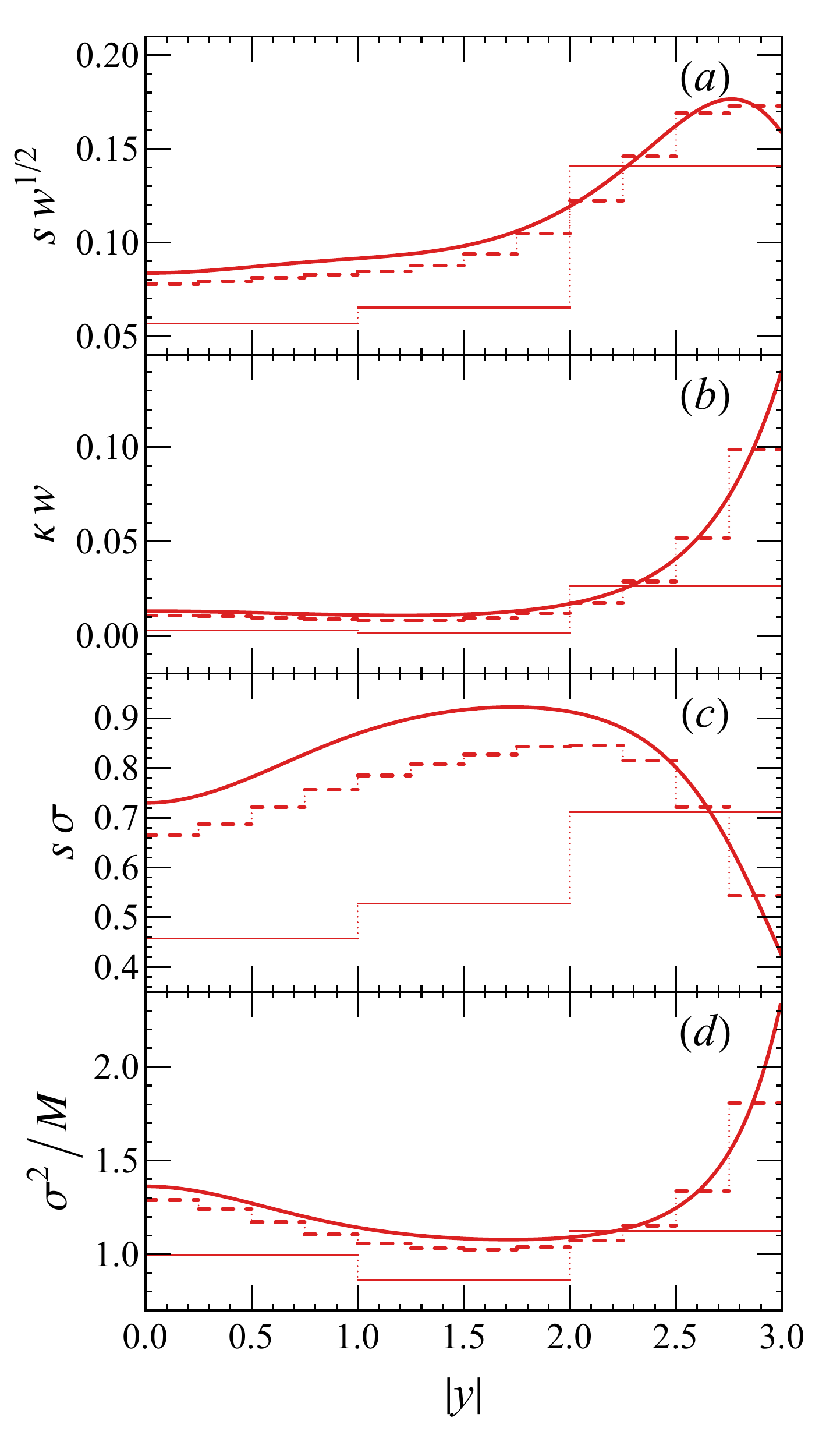}}
	\caption{Same as Figs. \ref{fig3} and \ref{fig4} but for (scaled) statistical measures: ($a$) $sw^{1/2}$, ($b$) $\kappa w$, ($c$) $s\sigma$ and ($d$) $\sigma^2/M$. Three different rapidity bins are selected: $w=0.01$ (solid), $w=0.25$ (dashed), and $w=1.00$ (thin solid).}
	\label{fig7}
\end{figure}

These cumulants allow for the calculation of statistical measures such as mean ($M$), variance ($\sigma^2$), skewness ($s$), and kurtosis ($\kappa$), which can be expressed as
\begin{align}
    M=\kappa_1^B\,,\quad
    \sigma^2=\kappa_2^B\,,\quad
    s=\frac{\kappa_3^B}{\left(\kappa_2^B\right)^{3/2}}\,,\quad
    \kappa=\frac{\kappa_4^B}{\left(\kappa_2^B\right)^{2}}\,.
\end{align}
Moreover, the product of these measures can be expressed in terms of the ratios of cumulants, as demonstrated below
\begin{align}
    \kappa\sigma^2=\frac{\kappa_4^B}{\kappa_2^B}\,,\quad
    s\sigma=\frac{\kappa_3^B}{\kappa_2^B}\,,\quad
    \frac{\sigma^2}{M}=\frac{\kappa_2^B}{\kappa_1^B}\,.
\end{align}
Figures \ref{fig3}($a$), \ref{fig3}($b$) and \ref{fig4}($b$) present $M/w$, $\sigma^2/w$, and $\kappa\sigma^2$, respectively. The other scaled statistical measures are displayed in Fig. \ref{fig7}.
\section{Cumulants of fine scan and GCE (Skellam distribution)}\label{AppD}
In the main text, we discussed that the system within the entire phase space is characterized by the CE since the total net-baryon number is conserved. However, experimental measurements are confined to finite rapidity windows, and baryons and antibaryons are subject to fluctuations. As the acceptance window decreases, the constraint from the baryon conservation weakens, allowing for greater independence in fluctuations of both baryons and antibaryons around their expected values~\cite{Koch:2008ia, Vovchenko:2020tsr}. In this appendix, we demonstrate the convergence of CE cumulants towards those derived from the Skellam distribution for the GCE when the rapidity window is sufficiently small.

\begin{figure}[!hbtp]
	\centering
	{\includegraphics[width=0.4\textwidth]{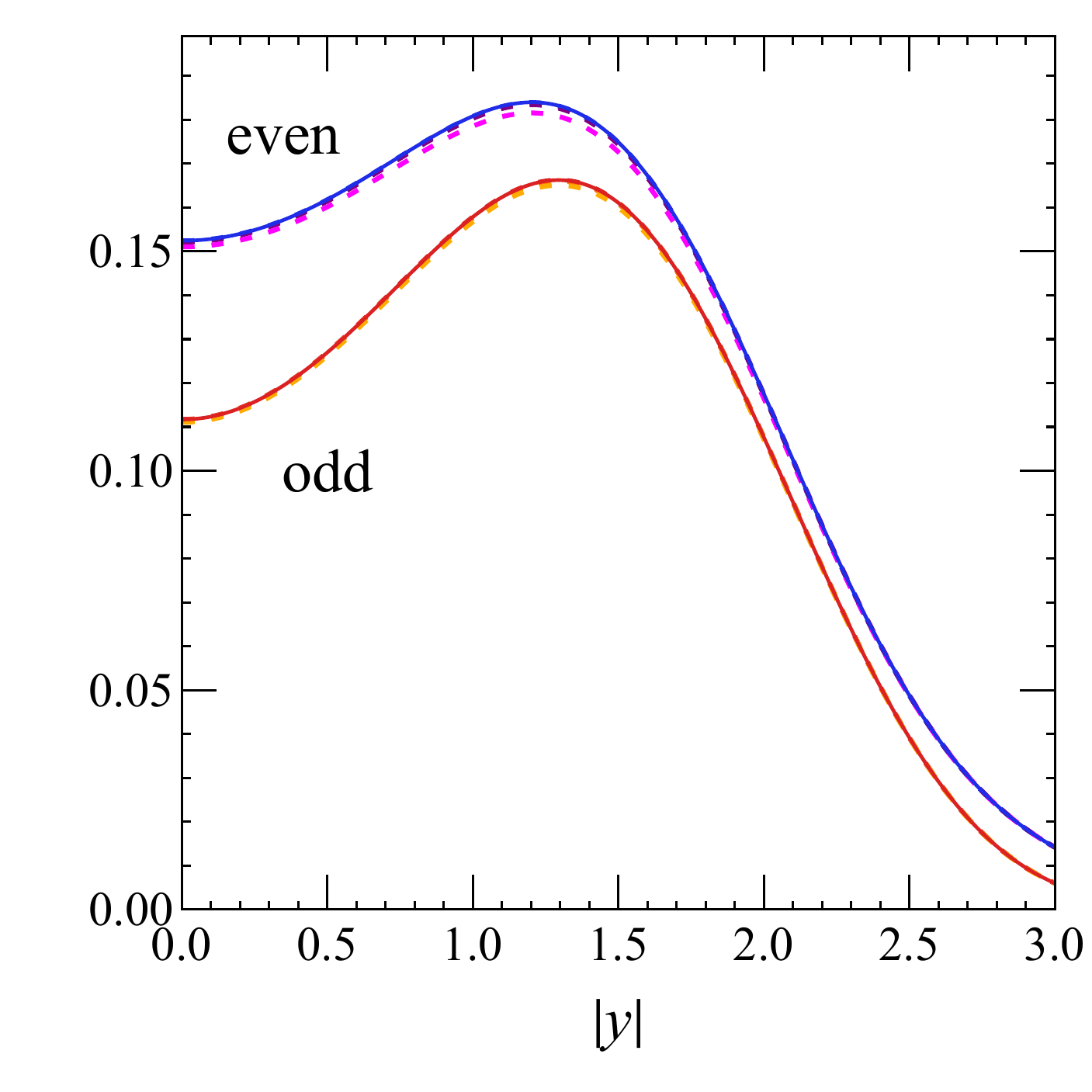}}
	\caption{
 Comparison of the first to sixth order cumulants obtained from a fine rapidity bin with $w=0.002$ (dashed) against the cumulants derived from the Skellam distribution for the GCE (solid), as given in Eq.~(\ref{statistical thermal cumulants}).
 }
	\label{Fig8}
\end{figure}
Besides deriving cumulants from generating functions, an alternative approach involves calculating them using susceptibilities denoted as $\chi_n^B$: $\kappa_n^B=VT^3\chi_n^B$. These susceptibilities are readily obtained from the thermodynamic pressure $P$ through the relationship $\chi_n^B=T^{n-4}\partial^{n}P/\partial\mu_B^n$. The thermodynamic pressure within the GCE of the statistical thermal model is
\begin{equation}
P^\mathrm{GCE}\left(T,V,\mu_B\right)=2\frac{T}{V}\cosh\left(\frac{\mu_B}{T}\right)z\left(T,V\right)\,,
\end{equation}
where we have adopted $B_h=1$ and taken the Maxwell--Boltzmann approximation. As a result, the net-baryon cumulants of both even and odd orders can be expressed as follows:
\begin{equation}
	\begin{gathered}
		\label{statistical thermal cumulants}
		\kappa_\mathrm{odd}^\mathrm{GCE}= 2\sinh\left(\frac{\mu_B}{T}\right)z\left(T,V\right)=\left<N_B\right>-\left<N_{\bar{B}}\right>\,,\\
		\kappa_\mathrm{even}^\mathrm{GCE}= 2\cosh\left(\frac{\mu_B}{T}\right)z\left(T,V\right)=\left<N_B\right>+\left<N_{\bar{B}}\right>\,.
	\end{gathered}
\end{equation}
From this, we can observe that $\kappa_\mathrm{odd}^\mathrm{GCE}/\kappa_\mathrm{odd}^\mathrm{GCE}=\kappa_\mathrm{even}^\mathrm{GCE}/\kappa_\mathrm{even}^\mathrm{GCE}=1$ and $\kappa_\mathrm{odd}^\mathrm{GCE}/\kappa_\mathrm{even}^\mathrm{GCE}=\tanh\left(\mu_B/T\right)$, which are also the results of the Skellam distribution~\cite{skellam1946frequency}. To demonstrate the convergence of net-baryon cumulants, when the rapidity window is small, toward the outcomes predicted by the Skellam distribution (i.e., the results of the GCE), we present a comparative analysis in FIG.~\ref{Fig8}. The figure effectively illustrates the consistency between the calculated first to sixth-order cumulants using the method in Sec.~\ref{Sec3} for a rapidity bin of $w=0.002$ and those computed using Eq. (\ref{statistical thermal cumulants}) for the GCE.
\section{Cumulant ratios with a symmetric rapidity window}\label{AppE}
\begin{figure}[!thbp]
	\centering
	{\includegraphics[width=0.4\textwidth]{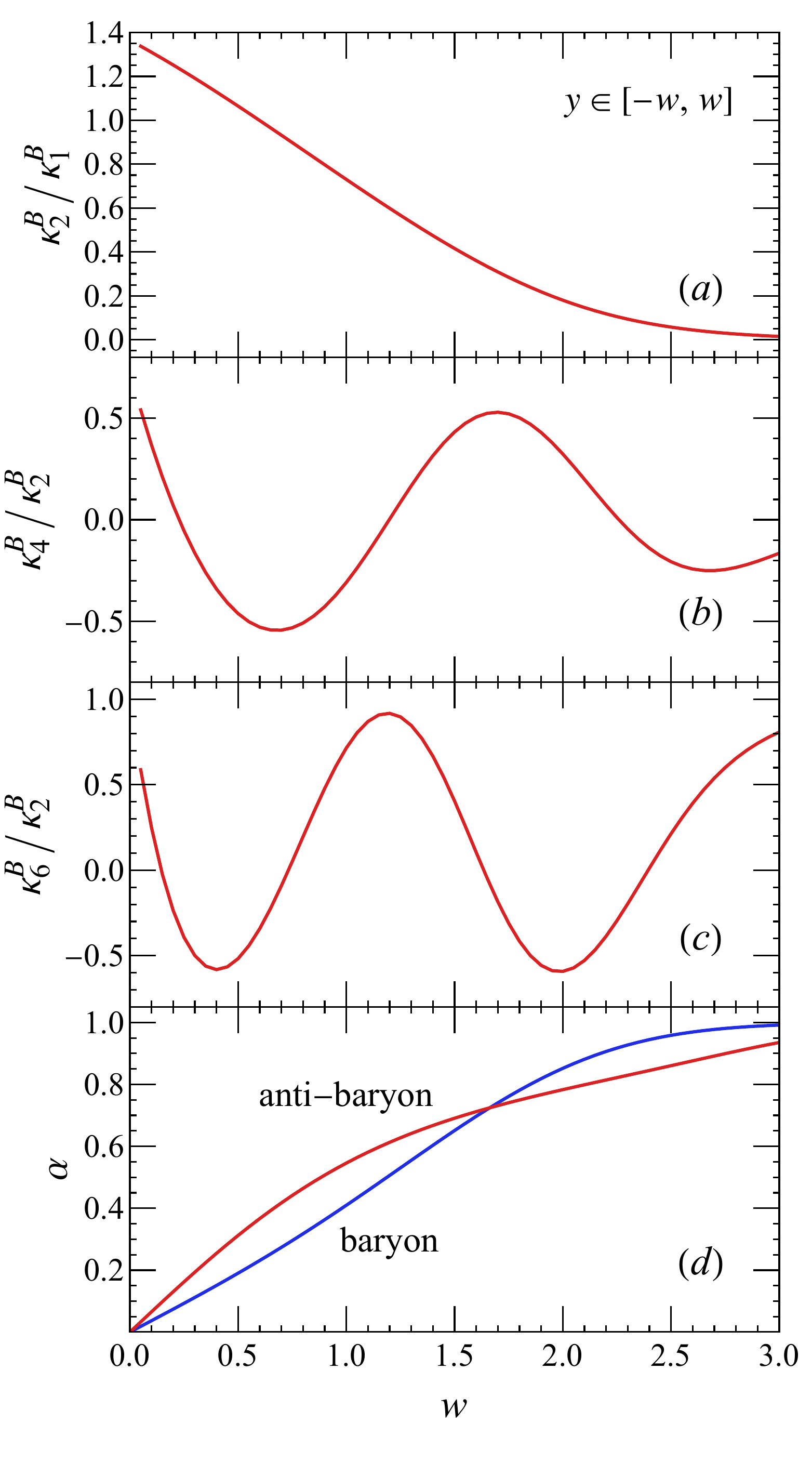}}
	\caption{
		Variation of the cumulant ratio  ($a$) $\kappa^B_2/\kappa^B_1$, ($b$) $\kappa^B_4/\kappa^B_2$, ($c$) $\kappa^B_6/\kappa^B_2$ and ($d$) baryon (blue) and antibaryon (red) acceptances as a function of the changing width of the rapidity window $y\in[-w,w]$.
	}
	\label{fig9}
\end{figure}
As a complement to FIG.~\ref{fig5}, we consider a similar scenario but with a symmetric rapidity window $y\in\left[-w,w\right]$. As illustrated in FIG.~\ref{fig9}, as the window broadens, and thus the acceptance increases, the high-order cumulant ratios exhibit more nonmonotonic behaviors compared to FIG.~\ref{fig5}.
\bibliography{Ref}
\end{document}